\documentclass[11pt,a4paper]{amsart}
\usepackage[T1]{fontenc}
\usepackage[utf8]{inputenc}
\usepackage{lmodern,amssymb,mathrsfs}
\usepackage{tikz}
\usetikzlibrary{arrows.meta}
\usepackage[hidelinks]{hyperref}
\hypersetup{pdftitle={Favard's Theorem on SU(2) and Quantum Computing},pdfauthor={Alexey Solyanik}}
\numberwithin{equation}{section}
\newtheorem{theorem}{Theorem}
\newtheorem{lemma}{Lemma}
\newtheorem*{corollary}{Corollary}
\newcommand{\R}{\mathbb R}
\newcommand{\C}{\mathbb C}
\newcommand{\sgn}{\operatorname{sgn}}

\newcommand{\norm}[1]{\left\|#1\right\|}
\newcommand{\SU}{\mathrm{SU}(2)}
\newcommand{\SO}{\mathrm{SO}(3)}
\title[Favard's theorem on SU(2) and quantum computing]{Favard's Theorem on \texorpdfstring{$\mathrm{SU}(2)$}{SU(2)} and Quantum Computing}
\author{Alexey Solyanik}
\address{Odesa Polytechnic National University, Odesa, Ukraine}
\email{solyanik@op.edu.ua}
\date{November 25, 2026}
\begin{document}
\raggedbottom
\begin{abstract}
We determine the sharp uniform error of linear approximation for the mean response of a quantum circuit as a single-qubit rotation varies. Using our sharp Favard-type estimates on \(S^3\) and the antipodal symmetry of their extremal function, we prove that for even responses on \(\mathrm{SU}(2)\) with \(\|-\Delta f\|_\infty\leq 1\), the optimal error using spins below \(k\) equals the full-sphere error for spherical harmonics of degrees below \(2k\). The H\"ormander--Bernhardsson extremal function gives the sharp constant in the \(k^{-2}\) bound. We also derive an exact expression for the largest error incurred when the average over all rotations is replaced by a weighted average over a prescribed finite set. As an application, we consider a circuit that uses the same gate twice. If only the fixed part of the circuit is compressed, the exact readout probabilities of the original and compressed circuits at the sixty rotational symmetries of an icosahedron determine the change in that probability at every rotation.
\end{abstract}
\maketitle
\setcounter{tocdepth}{1}
\tableofcontents

\section{Introduction}\label{sec:intro}

In this paper we apply methods of classical approximation theory to estimate the error of reconstructing a quantum circuit's response from measurements at finitely many settings. In the experiment considered here, the setting is the rotation performed by a gate, and the response is the probability of a specified outcome. We calculate this probability for the ideal quantum circuit. On a device, each run gives only one outcome, and only a finite set of randomly chosen rotations is tested. Then we use these measurements to fit a polynomial correction to the ideal response. The resulting bound on the error over the whole rotation group is expressed in terms of the degree and the number of runs and holds with prescribed probability.

A quantum computer is a computing device that stores information in the states of quantum systems and processes it by elementary controlled operations called gates. The idea arose in the early 1980s from the observation that the behaviour of quantum systems is hard to simulate on ordinary computers.

In 1982 Feynman proposed that one could use a controllable quantum system to simulate the dynamics of another~\cite[\S\,4]{FeynmanSimulation}. In 1985 Deutsch formulated a model of a universal programmable quantum computer~\cite[\S\,2]{DeutschUniversal}. In 1995 Monroe and coauthors realized a two-qubit controlled-NOT gate in a single trapped ion~\cite[p.~4714]{MonroeGate}. In a digital quantum simulation, a sequence of gates approximates the evolution of the system under study. Repeating a sequence from the same initial state allows one to estimate the probability of a chosen physical event. For example, Lanyon and coauthors used a register of trapped ions to follow how the probability that all four simulated spins point up changes with the simulated time~\cite[Fig.~4A]{LanyonSpin}. The measured quantity is thus a probability that depends on a parameter.

Information in a quantum computer is carried by qubits. Two internal energy levels of a trapped ion, for example, can serve as the basis states \(|0\rangle\) and \(|1\rangle\) of one qubit~\cite[p.~4714]{MonroeGate}. A pure state has the form \(\alpha|0\rangle+\beta|1\rangle\), where \(|\alpha|^2+|\beta|^2=1\). A measurement in this basis gives the two outcomes with probabilities \(|\alpha|^2\) and \(|\beta|^2\)~\cite{WatrousQubits}. The relative phase of the amplitudes matters once a gate acts on them. For example, the states \((|0\rangle+|1\rangle)/\sqrt{2}\) and \((|0\rangle-|1\rangle)/\sqrt{2}\) give the same probabilities in a direct measurement. The Hadamard gate carries \(|0\rangle\) and \(|1\rangle\) to these two states and is its own inverse. After it, therefore, they give outcomes \(0\) and \(1\), respectively, with certainty. A circuit arranges such operations in sequence so that interference affects the probability of the outcome one reads at the end~\cite[\S\S\,1.2--1.3]{NielsenChuang}.

Today quantum computers are built from trapped ions, superconducting circuits and other physical platforms. As of November 2026, the trapped-ion processor Helios operates with 98 qubits and has some of the most accurate gates reported to date~\cite{HeliosNature}.

We return to our problem. Here the parameter is the rotation performed by a gate, and even in a circuit with just one qubit the dependence of a probability on this rotation has a geometry of its own. For one qubit, each pure state is represented by a point on the unit two-sphere, the Bloch sphere. The basis states lie at its poles, and a gate rotates the sphere~\cite[\S\,4.2]{NielsenChuang}. Assign signs \(+1\) and \(-1\) to the two outcomes of a measurement along the polar axis. The north pole gives \(+1\) with certainty, the south pole gives \(-1\) with certainty, and either sign is equally likely on the equator. Now let \(R_y(\theta)\) and \(R_z(\theta)\) denote rotations about the indicated axes, with products acting from right to left. Both \(U=R_y(\pi/2)\) and \(V=R_y(\pi/2)R_z(\pi)\) carry the north pole \(e_z\) to \(e_x\) on the equator. On the second application \(U\) carries \(e_x\) to the south pole and \(V\) carries it to the north pole. Thus two gates that produce the same intermediate state give opposite final readings (Figure~\ref{fig:bloch}). The response to a repeated gate is a function of the rotation itself, not only of the state it prepares.

\begin{figure}[ht]
\centering
\begin{tikzpicture}[x=0.021cm,y=-0.021cm]
\draw[line width=0.5pt] (170,190) circle (110);
\draw[line width=0.4pt] (125,221) -- (134,222) -- (143,222) -- (153,223) -- (162,223) -- (172,223) -- (182,223) -- (191,223) -- (201,222) -- (210,221) -- (219,220) -- (228,219) -- (236,217) -- (243,215) -- (250,213) -- (256,211) -- (262,208) -- (267,206) -- (271,203) -- (275,200) -- (277,197) -- (279,195) -- (280,192);
\draw[line width=0.3pt,dash pattern=on 1.5pt off 2pt] (280,192) -- (280,189) -- (279,186) -- (277,183) -- (275,180) -- (271,177) -- (267,174) -- (262,172) -- (257,169) -- (251,167) -- (244,165) -- (236,163) -- (228,162) -- (220,160) -- (211,159) -- (201,158) -- (192,157) -- (182,157) -- (173,157) -- (163,157) -- (153,157) -- (144,158) -- (134,158) -- (125,159) -- (117,161) -- (108,162) -- (101,164) -- (93,166) -- (87,168) -- (81,171) -- (75,173) -- (71,176) -- (67,178) -- (64,181) -- (62,184) -- (60,187) -- (60,190);
\draw[line width=0.4pt] (60,190) -- (60,193) -- (62,196) -- (64,199) -- (67,202) -- (70,204) -- (75,207) -- (80,209) -- (86,212) -- (93,214) -- (100,216) -- (108,218) -- (116,219) -- (125,221);
\draw[line width=0.35pt] (170,85) -- (166,88) -- (162,92) -- (158,97) -- (154,102) -- (151,108) -- (147,115) -- (144,122) -- (141,130) -- (138,139) -- (135,147) -- (133,156) -- (131,166) -- (129,175) -- (127,185) -- (126,194) -- (125,204) -- (125,213) -- (125,223) -- (125,232) -- (126,241) -- (127,249) -- (128,257) -- (130,264) -- (132,271) -- (134,277) -- (136,283) -- (139,287) -- (142,291) -- (145,295) -- (149,297) -- (152,298);
\draw[line width=0.3pt,dash pattern=on 1.5pt off 2pt] (152,298) -- (156,299) -- (160,299) -- (164,298) -- (168,296) -- (172,293) -- (176,290) -- (180,286) -- (184,281) -- (188,275) -- (191,268) -- (195,261) -- (198,254) -- (201,246) -- (204,237) -- (206,228) -- (208,219) -- (210,210) -- (212,200) -- (213,190) -- (214,181) -- (215,171) -- (215,162) -- (215,153) -- (215,144) -- (214,135) -- (213,127) -- (211,119) -- (209,112) -- (207,106) -- (205,100) -- (202,95) -- (199,91) -- (196,87) -- (193,84) -- (189,82) -- (186,81);
\draw[line width=0.35pt] (186,81) -- (182,81) -- (178,81) -- (174,83) -- (170,85);
\draw[line width=0.5pt] (510,190) circle (110);
\draw[line width=0.4pt] (465,221) -- (474,222) -- (483,222) -- (493,223) -- (502,223) -- (512,223) -- (522,223) -- (531,223) -- (541,222) -- (550,221) -- (559,220) -- (568,219) -- (576,217) -- (583,215) -- (590,213) -- (596,211) -- (602,208) -- (607,206) -- (611,203) -- (615,200) -- (617,197) -- (619,195) -- (620,192);
\draw[line width=0.3pt,dash pattern=on 1.5pt off 2pt] (620,192) -- (620,189) -- (619,186) -- (617,183) -- (615,180) -- (611,177) -- (607,174) -- (602,172) -- (597,169) -- (591,167) -- (584,165) -- (576,163) -- (568,162) -- (560,160) -- (551,159) -- (541,158) -- (532,157) -- (522,157) -- (513,157) -- (503,157) -- (493,157) -- (484,158) -- (474,158) -- (465,159) -- (457,161) -- (448,162) -- (441,164) -- (433,166) -- (427,168) -- (421,171) -- (415,173) -- (411,176) -- (407,178) -- (404,181) -- (402,184) -- (400,187) -- (400,190);
\draw[line width=0.4pt] (400,190) -- (400,193) -- (402,196) -- (404,199) -- (407,202) -- (410,204) -- (415,207) -- (420,209) -- (426,212) -- (433,214) -- (440,216) -- (448,218) -- (456,219) -- (465,221);
\draw[line width=0.35pt] (510,85) -- (506,88) -- (502,92) -- (498,97) -- (494,102) -- (491,108) -- (487,115) -- (484,122) -- (481,130) -- (478,139) -- (475,147) -- (473,156) -- (471,166) -- (469,175) -- (467,185) -- (466,194) -- (465,204) -- (465,213) -- (465,223) -- (465,232) -- (466,241) -- (467,249) -- (468,257) -- (470,264) -- (472,271) -- (474,277) -- (476,283) -- (479,287) -- (482,291) -- (485,295) -- (489,297) -- (492,298);
\draw[line width=0.3pt,dash pattern=on 1.5pt off 2pt] (492,298) -- (496,299) -- (500,299) -- (504,298) -- (508,296) -- (512,293) -- (516,290) -- (520,286) -- (524,281) -- (528,275) -- (531,268) -- (535,261) -- (538,254) -- (541,246) -- (544,237) -- (546,228) -- (548,219) -- (550,210) -- (552,200) -- (553,190) -- (554,181) -- (555,171) -- (555,162) -- (555,153) -- (555,144) -- (554,135) -- (553,127) -- (551,119) -- (549,112) -- (547,106) -- (545,100) -- (542,95) -- (539,91) -- (536,87) -- (533,84) -- (529,82) -- (526,81);
\draw[line width=0.35pt] (526,81) -- (522,81) -- (518,81) -- (514,83) -- (510,85);
\draw[line width=0.5pt] (25,170) -- (29,171) -- (34,171) -- (39,172) -- (44,173) -- (49,173) -- (54,174) -- (59,175);
\draw[line width=0.35pt,dash pattern=on 1.5pt off 2pt] (59,175) -- (64,175) -- (69,176) -- (74,177) -- (79,178) -- (84,178) -- (89,179) -- (94,180) -- (99,180) -- (103,181) -- (108,182) -- (113,182) -- (118,183) -- (123,184) -- (128,184) -- (133,185) -- (138,186) -- (143,186) -- (148,187) -- (153,188) -- (158,188) -- (163,189) -- (168,190) -- (172,190) -- (177,191) -- (182,192) -- (187,192) -- (192,193) -- (197,194) -- (202,194) -- (207,195) -- (212,196) -- (217,196) -- (222,197) -- (227,198) -- (232,198) -- (237,199) -- (241,200) -- (246,200) -- (251,201) -- (256,202) -- (261,202) -- (266,203);
\draw[line width=0.5pt] (266,203) -- (271,204) -- (276,205) -- (281,205) -- (286,206) -- (291,207) -- (296,207) -- (301,208) -- (306,209) -- (311,209) -- (315,210);
\draw[line width=0.6pt,-{Stealth[length=4pt]}] (296.8,215.9) -- (297.1,217.6) -- (297.5,219.0) -- (298.1,220.2) -- (298.7,221.2) -- (299.4,221.8) -- (300.1,222.2) -- (301.0,222.3) -- (301.8,222.1) -- (302.6,221.6) -- (303.5,220.8) -- (304.3,219.7) -- (305.1,218.4) -- (305.8,216.9) -- (306.5,215.1) -- (307.1,213.3) -- (307.5,211.3) -- (307.9,209.3) -- (308.1,207.2) -- (308.3,205.2) -- (308.3,203.2) -- (308.1,201.3) -- (307.9,199.6) -- (307.5,198.0) -- (307.1,196.7) -- (306.5,195.6) -- (305.8,194.7) -- (305.1,194.2) -- (304.3,193.9) -- (303.5,194.0) -- (302.6,194.3) -- (301.8,195.0) -- (301.0,195.9) -- (300.1,197.1) -- (299.4,198.5) -- (298.7,200.1) -- (298.1,201.9) -- (297.5,203.8) -- (297.1,205.8) -- (296.8,207.9);
\draw[line width=0.5pt] (270,204) circle (1.4pt);
\draw[line width=0.8pt,-{Stealth[length=4pt]}] (119,232) -- (119,239) -- (120,246) -- (121,253) -- (121,259) -- (123,265) -- (124,271) -- (125,277) -- (127,282) -- (129,287) -- (131,292) -- (133,296) -- (135,300) -- (138,303) -- (140,306) -- (143,308) -- (145,310) -- (148,312) -- (151,313) -- (154,313) -- (157,313) -- (161,313) -- (164,312) -- (167,310);
\fill (170,85) circle (2.2pt);
\fill (125,221) circle (2.2pt);
\fill (170,295) circle (2.2pt);
\draw[line width=0.35pt] (478,137) -- (487,139) -- (496,141) -- (506,143) -- (515,145) -- (524,148) -- (533,152) -- (542,155) -- (551,159) -- (559,163) -- (567,167) -- (575,172) -- (582,176) -- (589,181) -- (594,186) -- (600,191) -- (604,195) -- (608,200) -- (611,205) -- (613,209) -- (615,214) -- (615,218) -- (615,222);
\draw[line width=0.3pt,dash pattern=on 1.5pt off 2pt] (615,222) -- (614,226) -- (612,229) -- (609,232) -- (606,235) -- (602,238) -- (597,240) -- (591,242) -- (585,243) -- (578,244) -- (571,244) -- (563,244) -- (555,244) -- (546,243) -- (537,242) -- (528,240) -- (519,238) -- (510,236) -- (501,233) -- (491,230) -- (482,227) -- (473,223) -- (465,219) -- (457,215) -- (449,210) -- (441,206) -- (435,201) -- (428,197) -- (423,192) -- (418,187) -- (414,182) -- (410,177) -- (408,173) -- (406,168) -- (405,164) -- (405,160) -- (405,156);
\draw[line width=0.35pt] (405,156) -- (407,152) -- (409,149) -- (412,146) -- (416,143) -- (421,141) -- (426,139) -- (432,138) -- (438,137) -- (445,136) -- (453,136) -- (461,136) -- (469,136) -- (478,137);
\draw[line width=0.35pt,dash pattern=on 1.5pt off 2pt] (556,266) -- (555,264) -- (553,261) -- (552,258) -- (550,256) -- (548,253) -- (547,251) -- (545,248) -- (544,246) -- (542,243) -- (541,240) -- (539,238) -- (537,235) -- (536,233) -- (534,230) -- (533,227) -- (531,225) -- (530,222) -- (528,220) -- (526,217) -- (525,215) -- (523,212) -- (522,209) -- (520,207) -- (519,204) -- (517,202) -- (515,199) -- (514,196) -- (512,194) -- (511,191) -- (509,189) -- (508,186) -- (506,184) -- (505,181) -- (503,178) -- (501,176) -- (500,173) -- (498,171) -- (497,168) -- (495,165) -- (494,163) -- (492,160) -- (490,158) -- (489,155) -- (487,153) -- (486,150) -- (484,147) -- (483,145) -- (481,142) -- (479,140);
\draw[line width=0.5pt] (479,140) -- (478,137) -- (476,134) -- (475,132) -- (473,129) -- (472,127) -- (470,124) -- (468,122) -- (467,119) -- (465,116) -- (464,114);
\draw[line width=0.6pt,-{Stealth[length=4pt]}] (482,119) -- (481,116) -- (480,113) -- (478,111) -- (475,109) -- (472,108) -- (469,108) -- (466,109) -- (463,110) -- (460,112) -- (457,114) -- (455,117) -- (454,120) -- (454,123) -- (455,126) -- (456,129) -- (458,131) -- (461,132) -- (464,133) -- (468,133) -- (471,132) -- (474,131) -- (477,129) -- (479,126);
\draw[line width=0.5pt] (478,137) circle (1.4pt);
\draw[line width=0.8pt,-{Stealth[length=4pt]}] (464,225) -- (472,225) -- (481,225) -- (490,223) -- (499,221) -- (507,218) -- (515,214) -- (523,210) -- (531,204) -- (538,199) -- (544,192) -- (550,186) -- (555,178) -- (560,171) -- (563,163) -- (566,155) -- (568,147) -- (569,139) -- (568,131) -- (568,123) -- (566,116) -- (563,109) -- (559,102) -- (555,96) -- (550,90) -- (544,85) -- (537,81) -- (530,77) -- (523,74) -- (515,72);
\fill (510,85) circle (2.2pt);
\fill (465,221) circle (2.2pt);
\fill (510,295) circle (2.2pt);
\node[right] at (178,68) {$+1$};
\node[right] at (180,306) {$-1$};
\node[left] at (120,236) {$e_x$};
\node[left] at (114,296) {$U$};
\node[left] at (502,68) {$+1$};
\node[right] at (520,306) {$-1$};
\node[left] at (460,236) {$e_x$};
\node[right] at (578,136) {$V$};
\node at (170,372) {$U=\dfrac{1}{\sqrt2}\begin{pmatrix}1&-1\\1&1\end{pmatrix}$};
\node at (170,414) {$\sqrt{Y}$ gate};
\node at (510,372) {$V=\dfrac{1}{\sqrt2}\begin{pmatrix}1&1\\1&-1\end{pmatrix}$};
\node at (510,414) {Hadamard gate};
\end{tikzpicture}
\caption{The Bloch sphere. At the north pole the reading is \(+1\) with probability one, and at the south pole it is \(-1\) with probability one. The gates \(U\) and \(V\) act on the sphere as rotations, that is, as elements of \(\mathrm{SO}(3)=\mathrm{SU}(2)/\{\pm I\}\). Since \(U\) and \(-U\) give the same rotation, a response that depends on the gate only through this rotation is an even function on \(\mathrm{SU}(2)\).}\label{fig:bloch}
\end{figure}

For each rotation we fix a pulse sequence and use it in both applications. This matters on a device because two pulse sequences implementing the same ideal gate can have different errors. Write \(t_2\) for the ideal mean of the final sign and \(f\) for its mean on the device. Since the probability of reading \(+1\) is \((1+f)/2\), half the absolute value of \(r=f-t_2\) is the error in that probability. The group \(\SU\) is isomorphic to the group of unit quaternions on \(S^3\). Opposite quaternions represent the same rotation, so \(r\) lifts to an even function on \(S^3\). On the unit sphere \(S^3\) we assume that \(r\) is continuous and has bounded Laplacian. A simple example is a dephasing channel between the two applications of the gate, for which this bound is computed explicitly (Section~\ref{sec:q-coefficient}). To approximate such a function we retain the harmonics with integer spins below \(k\), just as one retains the low frequencies of a periodic function. As far as we know, approximation on the group \(\SU\) was first studied by Ragozin~\cite{Ragozin1968}.

In 1937 Favard obtained the sharp error of approximation by one-dimensional harmonics on the circle, approximating the Bernoulli kernel in the integral norm~\cite[\S\,7, p.~219]{Favard1937}. We use the same approach. In our estimates on the sphere the Green kernel takes the place of the Bernoulli kernel~\cite{SolyanikLaplacian}. Spins below \(k\) on \(S^3\) correspond to spherical harmonics of degrees below \(2k\). For this even degree, the full-sphere problem admits an even extremal function, which therefore descends to the rotation group. We prove that, for a Laplacian bounded by one, the sharp linear error using spins below \(k\) equals \(h_k\), where \(h_k\) is the sharp constant of the Bohr--Favard type estimate~\cite{SolyanikLaplacian} on the whole sphere \(S^3\) at degree \(2k\) (Theorem~\ref{thm:even-sharp}). Its leading term, of order \(k^{-2}\), is determined by the H\"ormander--Bernhardsson extremal function. This tells us which degree suffices for a prescribed accuracy in the readout probability. We estimate separately the fluctuations due to the finite number of measurements (Section~\ref{sec:shots}). If the response depends only on the state prepared by the gate, as in data loading, it is a function on the Bloch sphere \(S^2\). For such responses the guarantee is about one quarter smaller and is given by an explicit series (Section~\ref{sec:hopf}).

Many procedures for testing a quantum processor replace the average over all gates by an average over a finite set. Such sets are called unitary \(t\)-designs. Examples are the Pauli group and the rotation groups of the tetrahedron, of the octahedron (the Clifford gates) and of the icosahedron. A design averages only polynomials exactly, while the response of a device is not a polynomial. We prove that, for a response with Laplacian bounded by one, the error of every \(t\)-design is at most \(h_{t+1}\), and we give an exact expression for the largest error. For independent designs on \(n\) qubits these errors simply add, and the number of qubits enters only through the number of terms (Section~\ref{sec:designs}).

A related problem arises when a circuit is shortened. The compressed circuit should give nearly the same probability of the outcome as the original one at every rotation \(U\). Its deviation from the original is a function on the group, and its mean over rotations is estimated by designs. More is possible when both circuits retain the same two uses of the gate \(U\), not controlled by other qubits, and only the remaining operations change. Then the difference of the responses is a polynomial of low degree, and no smoothness is needed. The exact probabilities for both circuits at the sixty rotational symmetries of an icosahedron determine their difference at every rotation (Section~\ref{sec:compression}).

Finally, for a gate used once, we consider the coefficient that measures how much of the ideal spin signal is present in the observed response. Given reversible access to numerical values of that response, we integrate its polynomial approximation classically and use a quantum algorithm to estimate the weighted mean of the remainder. For responses with bounded Laplacian we find the optimal orders of the error with \(N\) queries: \(N^{-2/3}\) for deterministic, \(N^{-7/6}\) for classical randomized and \(N^{-5/3}\) for quantum algorithms (Section~\ref{sec:q-optimal}).

\section{The group, the sphere and the Laplacian}\label{sec:geometry}

On the circle, Favard's argument rests on two facts: the exponentials diagonalize the second derivative, and the Bernoulli kernel inverts it on functions of mean zero. On the group of gates the spherical harmonics and the Green kernel of the Laplace--Beltrami operator play these roles. This section identifies a gate with a point of the sphere \(S^3\), translates the bound on the Laplacian into the language of rotation angles, and recalls the harmonics and the Green kernel that the rest of the paper uses.

Every element of \(\SU\) can be written uniquely as
\begin{equation}\label{eq:matrix}
 U(z,w)=\begin{pmatrix}z&-\bar w\\w&\bar z\end{pmatrix},
       \qquad |z|^2+|w|^2=1.
\end{equation}
The first column \((z,w)\) is a unit vector in \(\C^2\). The map \(U(z,w)\mapsto(z,w)\) gives a diffeomorphism from \(\SU\) onto the unit sphere \(S^3\subset\R^4\). Matrix multiplication on either side preserves this Euclidean norm, so both translations preserve the metric of the unit sphere and its normalized measure, which is the Haar probability. For unit vectors \(x,y\) representing two gates the real inner product equals \(\tfrac12\operatorname{Tr}(U(x)^*U(y))\), and their distance on the sphere is \(\arccos(x\cdot y)\).

The matrices \(U\) and \(-U\) act on every state in the same way, since they differ by a global phase. A physical gate is therefore a pair of antipodal points, and the gates form the quotient of \(S^3\) by \(x\mapsto-x\). This quotient is the rotation group \(\SO\) of the Bloch sphere, and the distance between two rotations is \(\arccos|x\cdot y|\). With this metric, a rotation of the three-dimensional Bloch vector through angle \(2\theta\), \(0\leq\theta\leq\pi/2\), lies at distance \(\theta\) from the identity on the quotient. This factor of two changes constants in estimates for a second derivative by a factor of four. Gr\"af gives the correspondence between the two polynomial systems and their cubatures~\cite[\S\S\,2--3]{Graf}; Beltr\'an and Ferizovi\'c use the rotation angle in their explicit Green formula~\cite[\S\,2.2]{BeltranFerizovic}.

The group law gives the smoothness condition of the introduction an interpretation in physical angles. Write \(\sigma_x,\sigma_y,\sigma_z\) for the Pauli matrices and \(R_\ell(\theta)=\exp(-\mathrm i\theta\sigma_\ell/2)\) for a pulse that rotates the Bloch vector about axis \(\ell\) through angle \(\theta\). For a smooth response,
\begin{equation}\label{eq:pulse-curvature}
 \Delta_{S^3}f(U)=
 4\sum_{\ell=x,y,z}\left.\frac{d^2}{d\theta^2}f(R_\ell(\theta)U)\right|_{\theta=0}.
\end{equation}
Indeed the three curves are great circles through \(U\). Their initial tangents are orthogonal and each has length \(1/2\). Taking the trace of the Hessian in the corresponding unit tangent directions gives the factor four. The hypothesis therefore bounds the sum of the three angular curvatures of the response as a function of the command. When the same gate is applied twice, as in the introduction, the perturbed command \(R_\ell(\theta)U\) replaces \(U\) in both applications.

Let \(\mathcal H_\ell\) be the restrictions to \(S^3\) of harmonic homogeneous polynomials of degree \(\ell\) on \(\R^4\). Writing the Euclidean Laplacian in polar coordinates and applying it to \(r^\ell Y(x)\) gives
\begin{equation}\label{eq:sphere-eigenvalue}
          -\Delta_{S^3}Y=\ell(\ell+2)Y,\qquad Y\in\mathcal H_\ell .
\end{equation}
Homogeneous harmonic decomposition gives \(\dim\mathcal H_\ell=(\ell+1)^2\), and the sum of the blocks through degree \(n\) is precisely the space of restrictions of polynomials of degree at most \(n\)~\cite[Chapter~IV, \S\,2]{SteinWeiss}. The sign of the block under \(x\mapsto-x\) is \((-1)^\ell\).

For the group interpretation, let \(\SU\) act on the \(\ell\)-th symmetric power of \(\C^2\). This irreducible unitary representation has dimension \(\ell+1\). Its character at a matrix with eigenvalues \(e^{i\theta},e^{-i\theta}\) is
\begin{equation}\label{eq:character-intro}
 \chi_\ell(\theta)=\sum_{r=0}^{\ell}e^{i(\ell-2r)\theta}
                  =\frac{\sin((\ell+1)\theta)}{\sin\theta}.
\end{equation}
The addition kernel of the spherical block is \((\ell+1)\chi_\ell(\arccos(x\cdot y))\) by the addition formula~\cite[Theorem~1.2.6]{DaiXu}. By Schur orthogonality, the projection kernel for the matrix coefficients of the symmetric power \(\pi_\ell\) is \((\ell+1)\operatorname{Tr}(\pi_\ell(U(y))^*\pi_\ell(U(x)))\), which equals the same character kernel. The equality of these kernels identifies the two spaces, with all matrix indices included. A single character describes the convolution kernel of an invariant projection. The functions acted on may depend on all three coordinates of the group.

In other words, the spherical harmonics of degree \(\ell\) on \(S^3\) are exactly the matrix coefficients of the representation \(\pi_\ell\), which physicists call spin \(j=\ell/2\). Harmonic analysis on the sphere and harmonic analysis on the group are the same thing. We use the two languages side by side: degree \(\ell=2j\) on the sphere corresponds to spin \(j\) on the group, and \(-\Delta_{S^3}\) has the eigenvalue \(\ell(\ell+2)=4j(j+1)\) on it. Since \(\pi_\ell(-U)=(-1)^\ell\pi_\ell(U)\), the even degrees, or integer spins, are the functions of the rotation alone. The odd degrees, or half-integer spins, change sign with the global phase.

We write \(\mathcal L=-\Delta_{S^3}\) for the Laplace--Beltrami operator on the unit sphere and use this notation throughout the paper. If \(\mathcal Lf=g\) is bounded, then \(\int g=0\). Conversely, every bounded mean-zero \(g\) has a unique continuous solution of mean zero. Indeed, convolution with the integrable zonal kernel
\begin{equation}\label{eq:kernel-geometry}
          G(\cos\theta)=\tfrac12(\pi-\theta)\cot\theta-\tfrac14
\end{equation}
produces it. In the language of gates, \(G(x\cdot y)\) is the response whose curvature is concentrated at the single command \(y\), balanced by a uniform curvature of the opposite sign elsewhere. Write \(h(\theta)=G(\cos\theta)\) for its angular profile. The measure in \(\theta\) is \((2/\pi)\sin^2\theta\,d\theta\). Away from the pole \((\sin^2\theta\,h'(\theta))'=\sin^2\theta\), and the negative flux \(-(2/\pi)\sin^2\theta\,h'(\theta)\) tends to 1 at zero. The constant in \eqref{eq:kernel-geometry} makes the integral zero. Integration by parts against a smooth test function therefore gives \(\mathcal LG=\delta_e-1\), for the pole \(e\). Continuity of \(G*g\) follows from continuity of translation in \(L^1\); uniqueness follows by taking the harmonic coefficients of a solution of \(\mathcal Lu=0\). This verifies the normalization of the full-sphere formula used in~\cite{SolyanikLaplacian}.

Isolated gates do not distinguish \(U\) from \(-U\), but circuits that control a gate by another qubit do. In the Hadamard test an auxiliary qubit prepared in \((|0\rangle+|1\rangle)/\sqrt2\) controls the gate \(U\) acting on a state \(|\psi\rangle\). The mean of a subsequent measurement of the auxiliary qubit in the same basis is \(\operatorname{Re}\langle\psi|U|\psi\rangle\). The global phase of \(U\) has become a relative phase between the two branches. This response changes sign under \(U\mapsto-U\). Phase estimation, the basis of many quantum algorithms, is built from the same controlled gates~\cite{CleveRevisited},~\cite[\S\,5.2]{NielsenChuang}. For such interferometric responses the relevant class consists of all functions on \(S^3\), with harmonics of every degree, and the constants of the full sphere are the sharp ones.

Without identifying antipodes, let \(H(n,S^3)\) be the supremum of \(\|f\|_\infty\) over functions with \(\|\mathcal Lf\|_\infty\leq1\) orthogonal to all harmonics of degrees \(0,\ldots,n-1\). The letter \(H\) refers to Harald Bohr, who first estimated functions with missing low harmonics in this way on the circle. In our work~\cite{SolyanikLaplacian} it is shown that for the full group
\begin{equation}\label{eq:full-favard}
 H(n,S^3)=\min_{\deg p<n}\int_{-1}^1 |G(t)-p(t)|\,d\mu(t),
 \qquad d\mu(t)=\tfrac2\pi\sqrt{1-t^2}\,dt .
\end{equation}
For arbitrary \(f\) with bounded Laplacian, if \(p\) is the minimizing polynomial, the approximant with value \(\bar f+\int p(x\cdot y)\mathcal Lf(y)\,d\sigma(y)\) has error at most \(H(n,S^3)\|\mathcal Lf\|_\infty\). Integration by parts moves \(\mathcal L\) to \(p\), so this expression depends only on the part of \(f\) of degree below \(n\). The associated finite-degree bound and limit are
\begin{equation}\label{eq:full-error-bound}
 H(n,S^3)\leq\frac{\beta_3}{n^2},\qquad
 \lim_{n\to\infty}n^2H(n,S^3)=\beta_3 ,
 \qquad \beta_3\simeq 2.850999422 .
\end{equation}
Here, if \(\varphi\) is the integrable entire function of exponential type at most one minimizing \(\|\varphi\|_1\) subject to \(\varphi(0)=1\), then~\cite{SolyanikLaplacian}
\begin{equation}\label{eq:beta-definition}
 \beta_3=\frac{\|\varphi\|_1}{2\pi}\int_0^\infty\frac{dy}{\varphi(iy)} .
\end{equation}

The entire function in this expression is the one-dimensional extremal introduced by H\"ormander and Bernhardsson in their extension of Bohr's inequality~\cite{HormanderBernhardsson}. Bondarenko, Ortega-Cerdà, Radchenko and Seip~\cite[Theorem~1.1 and formula~(1.4)]{BondarenkoHB} characterize a factor of the extremal by a second-order differential equation and express its point-evaluation constant through its zeros. Their exponential type is \(\pi\). If \(\phi_\pi\) denotes their extremal with \(\phi_\pi(0)=1\), then \(\varphi(z)=\phi_\pi(z/\pi)\) in the expression for \(\beta_3\).

For the numerical choice of a degree it suffices to know that \(\beta_3<3\).

The interpolation behind \eqref{eq:full-favard} can be seen directly from the kernel. With \(-1<t<1\), the substitution \(s=\tanh(v/2)\) gives
\begin{equation}\label{eq:green-integral}
 G(t)+\tfrac14=\frac12\int_0^\infty\frac{t}{\cosh v-t}\,dv .
\end{equation}
Every positive-order derivative of this expression is strictly positive. For example its \(a\)-th derivative is the integral of
\((a!/2)\cosh v\,(\cosh v-t)^{-a-1}\).
Let \(Q_n\) be a monic polynomial of degree \(n\) of least integral norm for \(d\mu\). Its sign annihilates every lower degree by first variation. It has \(n\) distinct interior zeros, since otherwise its sign could be matched by a lower-degree polynomial. Interpolating \(G\) at these zeros gives a remainder of sign \(\sgn Q_n\), by the positive derivative. Pairing with this sign proves minimality in \eqref{eq:full-favard}. The same sign, viewed as a zonal function on the sphere, is a bounded source for an extremal potential.

\section{Sharp estimates of linear approximation for even functions}\label{sec:even-sharp}

In our previous work~\cite{SolyanikLaplacian} the sharp estimate of linear approximation on \(S^3\) was found for functions whose Laplacian is bounded by one, by polynomials in spherical harmonics of degree strictly less than \(n\). This is the constant \(H(n,S^3)\) in~\eqref{eq:full-favard}. A gate is determined by a rotation, and \(U\) and \(-U\) give the same rotation, so gate responses are even functions on \(S^3\), and only the even harmonics of degrees \(0,2,\ldots,2k-2\), that is, the spins below \(k\), are available. In this section we find the sharp estimate of linear approximation for even functions. One could expect it to be smaller than on the whole sphere. It is not: it equals \(H(2k,S^3)\) (Theorem~\ref{thm:even-sharp}), and our estimates on \(S^3\) bound it by \(\beta_3/(4k^2)\) with a constant that cannot be decreased.

At cutoff \(k\) the approximating polynomial uses spins \(0,\ldots,k-1\). Spin zero is the average over all commands. Spin one describes, for example, the one-use mean \(R_{33}\) of Section~\ref{sec:q-coefficient}. The two-use mean already contains spin two.

Throughout this section \(d\sigma\) is normalized area on the unit sphere
\(S^3\), and \(\mathcal L=-\Delta_{S^3}\) as in Section~\ref{sec:geometry}. Write \(\mathcal H_\ell\) for its
eigenspace with eigenvalue \(\ell(\ell+2)\), \(C_+\) for the continuous
even functions, and put
\begin{equation}\label{eq:spin-space-and-class}
 \mathcal{V}_k=\bigoplus_{j=0}^{k-1}\mathcal{H}_{2j},\qquad
 \mathcal{B}^+=\{ f\in C_+ : \|\mathcal Lf\|_\infty\leq1\}.
\end{equation}
The functions may be
real or complex. Since the class contains arbitrary constants, any
linear method with finite error must reproduce constants.

We now construct the optimal operator. It needs three objects: the Green kernel for even functions, the measure in which this kernel is approximated, and the nodes at which it is interpolated.

For an even function the Green kernel \(G\) of~\eqref{eq:kernel-geometry} may be replaced by its even part, which depends only on \(u=(x\cdot y)^2\):
\begin{equation}\label{eq:even-kernel}
 K(u)=\frac{G(\sqrt u)+G(-\sqrt u)}2
      =\frac{\sqrt u\,\arcsin\sqrt u}{2\sqrt{1-u}}-\frac14,
 \qquad 0\leq u<1.
\end{equation}
For a fixed \(x\) the variable \(u=(x\cdot y)^2\) has the probability law
\begin{equation}\label{eq:even-measure}
 d\nu(u)=\frac2\pi\sqrt{\frac{1-u}{u}}\,du,\qquad 0<u<1,
\end{equation}
and every \(f\in\mathcal B^+\) is recovered from its Laplacian and its mean:
\begin{equation}\label{eq:even-potential}
 f(x)-\bar f=\int_{S^3}K((x\cdot y)^2)\mathcal Lf(y)\,d\sigma(y),
 \qquad \bar f=\int_{S^3}f\,d\sigma .
\end{equation}
Indeed, the right side has the same Laplacian and the same mean as \(f-\bar f\).

For each integer \(k\ge1\) let \(q_k\) be the real monic polynomial of degree \(k\) with the least value of \(\int_0^1|q|\,d\nu\), and put \(s_k=\operatorname{sgn}q_k\). Its zeros \(0<u_1<\cdots<u_k<1\) are simple (see the proof below). Let \(p_{k-1}\) be the polynomial of degree less than \(k\) that interpolates \(K\) at these zeros, and put
\begin{equation}\label{eq:even-error}
 h_k=\int_0^1 |K-p_{k-1}|\,d\nu .
\end{equation}
The operator
\begin{equation}\label{eq:even-method}
 T_k f(x)=\bar f+\int_{S^3}
     p_{k-1}((x\cdot y)^2)\mathcal Lf(y)\,d\sigma(y)
\end{equation}
is linear and takes every even function to an even polynomial of degree less than \(2k\).

\begin{theorem}\label{thm:even-sharp}
For every integer \(k\ge1\), among all linear operators taking even functions to even polynomials of degree less than \(2k\), the operator \(T_k\) gives the best approximation of even functions \(f\) with \(\|\mathcal Lf\|_\infty\le1\), and
\begin{equation}\label{eq:even-equality}
 \sup_{\|\mathcal Lf\|_\infty\le1}\|f-T_kf\|_\infty=h_k=H(2k,S^3).
\end{equation}
\end{theorem}

\begin{proof}
We prove three facts: the constant \(h_k\) equals \(H(2k,S^3)\), the operator \(T_k\) has error at most \(h_k\), and no linear operator of this kind has a smaller error.

Symmetrizing a monic polynomial in \(t\) preserves its leading term and cannot increase its integral norm for the even weight \(d\mu\) in~\eqref{eq:full-favard}, and \(\int_{-1}^1|q(t^2)|\,d\mu(t)=\int_0^1|q|\,d\nu\). Hence \(q_k(t^2)\) is the monic polynomial of degree \(2k\) of least integral norm for \(d\mu\). By our work~\cite{SolyanikLaplacian} it is unique and has \(2k\) simple zeros in \((-1,1)\), so the zeros of \(q_k\) are simple, and minimality gives
\begin{equation}\label{eq:even-moments}
 \int_0^1 s_k(u)v(u)\,d\nu(u)=0,\qquad \deg v<k .
\end{equation}
If \(P\) interpolates \(G\) at the zeros \(\pm\sqrt{u_j}\) of \(q_k(t^2)\), its even part interpolates \(K(t^2)\) there and is therefore \(p_{k-1}(t^2)\). By our work, \(H(2k,S^3)=\int_{-1}^1 G(t)s_k(t^2)\,d\mu(t)=\int_{-1}^1|G-P|\,d\mu\). For every polynomial \(p\) of degree less than \(k\), \eqref{eq:even-moments} and symmetrization of \(G-P\) give
\[
 \int_0^1|K-p|\,d\nu\ \ge\int_0^1 K s_k\,d\nu
 =H(2k,S^3)=\int_{-1}^1|G-P|\,d\mu\ \ge h_k .
\]
Taking \(p=p_{k-1}\) shows that all these quantities are equal:
\begin{equation}\label{eq:even-min}
 h_k=\int_0^1 K s_k\,d\nu=\min_{\deg p<k}\int_0^1 |K-p|\,d\nu=H(2k,S^3).
\end{equation}
The minimum is the same over complex polynomials, since taking the real part can only decrease the pointwise error.

Since the polynomial kernel is smooth, the integral in~\eqref{eq:even-method} equals the integral of \(f(y)\) times \(\mathcal L_y p_{k-1}((x\cdot y)^2)\), so \(T_k\) is bounded on continuous even functions and reproduces constants. Subtracting~\eqref{eq:even-method} from~\eqref{eq:even-potential} bounds its error by \(h_k\|\mathcal Lf\|_\infty\).

It remains to construct an example on which no operator of this kind does better. The extremal function of our work for degree \(2k\) and pole \(e\) is the potential \(f_e\) of the source \(s_k((e\cdot y)^2)\). This source is even, so \(f_e\) is even, has \(\|\mathcal Lf_e\|_\infty=1\), has no harmonics of degree below \(2k\), and \(\|f_e\|_\infty=h_k\). Let \(A\) be any competitor. Averaging \(R^{-1}AR\) over the rotations \(R\) of \(S^3\) gives an operator with no larger error that commutes with rotations, hence with \(\mathcal L\). It annihilates every harmonic of degree at least \(2k\), and therefore \(f_e\), which is a uniform limit of even polynomials without harmonics of degree below \(2k\). Its error at \(f_e\) is \(h_k\). Averaging \(f_e\) against smooth nonnegative approximate identities on the rotation group shows that smooth functions come arbitrarily close to this error.
\end{proof}

The inverse-square law for gates follows at once from the full sphere, and its sharp coefficient comes from the H\"ormander--Bernhardsson extremal function.

\begin{corollary}
For every integer \(k\ge1\),
\begin{equation}\label{eq:spin-error-bound}
 4k^2h_k\le\beta_3<3,
\end{equation}
and the coefficient \(\beta_3\) cannot be decreased.
\end{corollary}

\begin{proof}
By Theorem~\ref{thm:even-sharp}, \(h_k=H(2k,S^3)\). Take \(n=2k\) in~\eqref{eq:full-error-bound}, which is our estimate~\cite{SolyanikLaplacian}. The same estimate gives \(4k^2h_k\to\beta_3\), so no smaller coefficient holds for all \(k\).
\end{proof}

For small \(k\) the bound~\eqref{eq:spin-error-bound} can be improved: \(h_k\) is computed exactly from the zeros of \(q_k\). Write them as \(u_j=\sin^2a_j\), \(0<a_1<\cdots<a_k<\pi/2\). For \(u=\sin^2a\) one has \(d\nu=(4/\pi)\cos^2a\,da\), \(K=\tfrac12a\tan a-\tfrac14\) and
\begin{equation}\label{eq:even-primitive}
 \int_0^{\sin^2a}K(u)\,d\nu(u)=-\frac{a\cos^2a}{\pi},
\end{equation}
so integration by parts across the jumps of \(s_k\) gives
\begin{equation}\label{eq:even-finite}
 h_k=\frac2\pi\sum_{j=1}^k(-1)^{k-j}a_j\cos^2 a_j ,
\end{equation}
where the nodes are the unique ordered solution of
\begin{equation}\label{eq:moment-node-equations}
 \int_0^{\pi/2}
  \operatorname{sgn}\!\left(\prod_{j=1}^k(\sin^2a-\sin^2a_j)\right)
  \sin^{2r}a\,\cos^2a\,da=0,\qquad 0\le r<k .
\end{equation}
For \(k=1\) the node is the median of \(\nu\), and
\begin{equation}\label{eq:even-first}
 2a_1+\sin(2a_1)=\frac{\pi}{2},\qquad
 h_1=\frac{2a_1\cos^2a_1}{\pi}\simeq 0.2215376104 .
\end{equation}
Since \(T_1f=\bar f\), this is the largest error in replacing an even function of bounded Laplacian by its mean.

Table~\ref{tab:cutoffs} answers a practical question: how many harmonics are needed for a prescribed accuracy? Suppose that the Laplacian of a response is bounded by \(S\), and we want to approximate the response with error at most \(\varepsilon\). By Theorem~\ref{thm:even-sharp} the operator \(T_k\) guarantees the error \(Sh_k\), so it suffices to take the smallest \(k\) with \(Sh_k\le\varepsilon\). The output of \(T_k\) is a polynomial of spins below \(k\), and it is described by \(\dim\mathcal V_k\) real coefficients, given in the third column. For example, for \(\varepsilon=0.03\,S\) the table gives \(k=4\), that is, \(84\) coefficients. Note that the estimate~\eqref{eq:spin-error-bound} requires \(k=5\) for the same \(\varepsilon\), which leads to \(165\) coefficients.

\begin{table}[ht]
\centering
\caption{The sharp linear error at each spin cutoff on the unit class, rounded to five decimal places. The values \(h_1,\ldots,h_5\) are certified in Appendix~\ref{app:computation}; \(h_6\) and \(h_7\) were computed from~\eqref{eq:moment-node-equations} in high-precision arithmetic.}\label{tab:cutoffs}
\begin{tabular}{@{}rlrr@{}}
\hline
\(k\) & Output spins & \(\dim\mathcal V_k\) & \(h_k\)\\
\hline
1 & \(0\) & 1 & 0.22154\\
2 & \(0,1\) & 10 & 0.08647\\
3 & \(0,1,2\) & 35 & 0.04673\\
4 & \(0,\ldots,3\) & 84 & 0.02939\\
5 & \(0,\ldots,4\) & 165 & 0.02022\\
6 & \(0,\ldots,5\) & 286 & 0.01477\\
7 & \(0,\ldots,6\) & 455 & 0.01127\\
\hline
\end{tabular}
\end{table}

For large \(k\) the table is not needed. The dimension of \(\mathcal V_k\) is \(k(2k-1)(2k+1)/3\approx\tfrac43k^3\), and by the Corollary it suffices that \(\beta_3S/(4k^2)\le\varepsilon\). Hence the number of coefficients that guarantees accuracy \(\varepsilon\) grows like
\begin{equation}\label{eq:parameter-budget}
 \dim\mathcal V_k\approx\tfrac16\Bigl(\frac{\beta_3S}{\varepsilon}\Bigr)^{3/2}.
\end{equation}
This is the number of parameters the optimal linear method needs to store, load or compress a response with accuracy \(\varepsilon\).

\section{Estimating a response from measurements}\label{sec:shots}

So far the response was a known function. For a real device it is not known: it has to be measured, and each measurement gives only a random sign \(\pm1\). Fix the preparation, the readout axis and the pulse recipe of each gate. Let \(f(U)\) be the mean of the sign read after two applications of the gate \(U\), let \(t_2(U)\) be the same mean for an ideal gate, and let \(r=f-t_2\) be the error of the device; the error in the probability of a positive readout is \(\tfrac12|r(U)|\). We want to find \(r\) at every gate from a finite number of measurements made at randomly chosen gates. The operator \(T_k\) is suited to this task, because it averages the response against a kernel, and an average can be estimated by random sampling. The bound~\eqref{eq:shot-map-uniform} below says how accurate this estimate is: its error is the approximation error \(S_rh_k\) of Theorem~\ref{thm:even-sharp} plus a random part that decreases like \(1/\sqrt m\) with the number \(m\) of measurements.

With \(u=(\tfrac12\operatorname{Tr}(U^*V))^2\), integration by parts in~\eqref{eq:even-method} gives
\begin{equation}\label{eq:shot-map-kernel}
 T_kf(U)=\int W_k(U,V)f(V)\,d\sigma(V),\qquad
 W_k(U,V)=1+\mathcal L_V\big[p_{k-1}(u)\big]=w_k(u),
\end{equation}
where \(w_k\) is given explicitly by~\eqref{eq:polynomial-density}; put \(B_k=\max_{0\le u\le1}|w_k(u)|\). Draw independent commands \(V_1,\ldots,V_m\) from Haar probability, run the circuit once at each and record the final sign \(X_i\). Assume that the trials are independent and \(\mathbb E[X_i\mid V_i]=f(V_i)\). Set
\begin{equation}\label{eq:shot-map-estimator}
 \widehat r_k(U)=\frac1m\sum_{i=1}^m W_k(U,V_i)X_i -T_k t_2(U),
 \qquad \widehat f_k(U)=t_2(U)+\widehat r_k(U).
\end{equation}
Each trial supplies one sign, whatever the number of prediction points. Put
\begin{equation}\label{eq:shot-bernstein-scale}
 v_k=\int_0^1 w_k(u)^2\,d\nu(u),\qquad C_k=B_k+\sqrt{v_k},\qquad
 b_m(z)=\frac{C_kz}{3m}+\sqrt{\left(\frac{C_kz}{3m}\right)^2+\frac{2v_kz}{m}} ,
\end{equation}
where \(\nu\) is the measure~\eqref{eq:even-measure}, and let \(L=2k-2\).

Suppose that \(r\) is continuous and even, with \(\|\mathcal Lr\|_\infty\le S_r\). Conditioning on \(V_i\) gives \(\mathbb E[W_k(U,V_i)X_i]=T_kf(U)\), so the estimator is unbiased for \(T_kr\), and by Theorem~\ref{thm:even-sharp} its bias is at most \(S_rh_k\). At a fixed command each centered summand is bounded by \(C_k\) and has variance at most \(v_k\), since \(X_i^2=1\) and Haar probability is invariant. Bernstein's inequality bounds the probability of a deviation \(s\) by \(2\exp[-ms^2/(2(v_k+C_ks/3))]\). With a union bound over a prescribed finite list \(F\) of commands this gives, for \(0<\alpha<1\) and with probability at least \(1-\alpha\),
\begin{equation}\label{eq:shot-map-list}
 \max_{U\in F}|\widehat r_k(U)-r(U)|\le S_rh_k+b_m\!\left(\log(2|F|/\alpha)\right).
\end{equation}

A bound on the whole group needs one more step, since there are infinitely many commands. The fluctuation \(P=\widehat r_k-T_kr\) is an even polynomial of degree at most \(L\), and such a polynomial cannot be large only at a single point. Choose \(0<\rho<\sqrt{2}/L\), and put \(\gamma=1-L^2\rho^2/2\) and \(\omega=(2\rho-\sin(2\rho))/\pi\). On every great circle through a point where \(|P|\) is maximal, Bernstein's inequality for trigonometric polynomials~\cite[p.~357]{Riesz1914} and Taylor's formula give \(|P|\ge\gamma\|P\|_\infty\) on the ball of radius \(\rho\) about that point, whose Haar probability is \(\omega\). Hence \(\Pr(\|P\|_\infty\ge s/\gamma)\le\omega^{-1}\,\mathbb E\,\sigma\{|P|\ge s\}\le(2/\omega)\exp[-ms^2/(2(v_k+C_ks/3))]\). Adding the bias, we obtain with probability at least \(1-\alpha\)
\begin{equation}\label{eq:shot-map-uniform}
 \|\widehat f_k-f\|_\infty=\|\widehat r_k-r\|_\infty
 \le S_rh_k+\frac{1}{\gamma}\,b_m\!\left(\log\frac{2}{\alpha\omega}\right)=:E_k .
\end{equation}
For \(k=1\) the fluctuation is constant, and one takes \(\omega=\gamma=1\).

What does an experiment finally report about the device? Usually not the whole function \(\widehat r_k\), but one number: the largest error in the probability of a positive readout over all commands, \(\mathcal P=\|r\|_\infty/2\). The bound~\eqref{eq:shot-map-uniform} turns the measurements into a confidence interval for this number. Indeed, \(\widehat r_k\) is a known polynomial, so its maximum can be enclosed numerically, \(a\le\|\widehat r_k\|_\infty\le b\). Then, with probability at least \(1-\alpha\),
\begin{equation}\label{eq:shot-map-discrepancy}
 \max\{0,(a-E_k)/2\}\le \mathcal P\le \min\{1,(b+E_k)/2\}.
\end{equation}
Appendix~\ref{app:computation} gives the mesh error of this maximization. Confidence bands for several cutoffs on the same data require dividing the failure probability among them.

How large should \(k\) be for a given number of measurements? A larger \(k\) decreases the approximation error but increases the random error, since the kernel \(W_k\) grows. The balance is seen best in the mean-square error over all commands, which splits exactly into these two parts:
\begin{equation}\label{eq:shot-map-risk}
 \mathbb E\|\widehat f_k-f\|_2^2
  =\|r-T_k r\|_2^2+
    \frac{v_k-\|T_k f\|_2^2}{m},
\end{equation}
since \(\int W_k(U,V)^2\,d\sigma(V)=v_k\) for every \(U\). The first term is the squared approximation error, and the second is the cost in measurements of retaining more spins. A certified polynomial \(P\) from Appendix~\ref{app:computation} may be used in place of \(p_{k-1}\), with \(\int_0^1|K-P|\,d\nu\) in place of \(h_k\).

\section{Averages over finite sets of gates}\label{sec:designs}

Many procedures for testing a quantum processor need an average over all gates. Randomized benchmarking applies long sequences of random gates and reads off an average error per gate from the decay of the survival probability~\cite{MagesanRB}. Classical shadows rotate an unknown state by random gates before measuring it and predict many of its properties from the recorded outcomes~\cite{HuangKuengPreskill}. Randomized compiling inserts random Pauli gates to turn coherent errors into stochastic ones~\cite{WallmanEmerson}. No experiment can draw from the Haar measure exactly. In practice the average is taken over a small finite set: the four Pauli gates, the twenty-four Clifford gates, or products of such sets on separate qubits.

This replacement is exact for the quantities these procedures were designed for. Such a quantity is a polynomial of low degree in the matrix entries of the gate and of its complex conjugate. A finite set with this exactness property is called a unitary design. For a device the situation is different. The commanded gates may form a design, while the response of the device to a command is not a polynomial. Suppose that a laboratory reports the mean of a measured response over the Clifford commands as its mean over all rotations. How large can the error be, and how strong a design is needed for a prescribed accuracy? Favard's approach answers both questions with the constants already found.

Let \(U_1,\ldots,U_N\in\SU\) and let \(w_1,\ldots,w_N\) be positive weights with sum one. Following Dankert, Cleve, Emerson and Livine~\cite{DankertDesigns} and Gross, Audenaert and Eisert~\cite{GrossAudenaertEisert}, the weighted set is a \emph{unitary \(t\)-design} if
\begin{equation}\label{eq:design-definition}
 \sum_{i=1}^N w_i\,U_i^{\otimes t}\otimes\overline{U_i}^{\otimes t}
 =\int_{\SU}U^{\otimes t}\otimes\overline U^{\otimes t}\,d\sigma(U).
\end{equation}
The usual designs have equal weights \(1/N\). Replacing \(U_i\) by \(-U_i\) does not change the left side, so a design is really a set of rotations. Every set is a \(0\)-design. We write \(Qf=\sum_iw_if(U_i)\) for the corresponding average.

Before estimating any error we translate this definition into the language of harmonics. The next lemma says that a design is exactly a set of gates whose average is correct for every polynomial of low spin.

\begin{lemma}\label{lem:design-cubature}
A weighted set is a unitary \(t\)-design if and only if \(Qp=\int p\,d\sigma\) for every \(p\in\mathcal V_{t+1}\), that is, for every polynomial of integer spin at most \(t\).
\end{lemma}

\begin{proof}
By~\eqref{eq:matrix} the entries of \(U\) are \(z,-\bar w,w,\bar z\), and those of \(\overline U\) are their conjugates. An entry of \(U^{\otimes t}\otimes\overline U^{\otimes t}\) is therefore a product of \(2t\) letters from \(z,\bar z,w,\bar w\), up to sign, and every such product occurs. These products span the homogeneous polynomials of degree \(2t\) in the four real coordinates. On the sphere, multiplication by \(|x|^2=1\) raises every lower even degree to \(2t\). Their restrictions are therefore exactly the even polynomials of degree at most \(2t\), which form \(\mathcal V_{t+1}=\bigoplus_{j\le t}\mathcal H_{2j}\) by Section~\ref{sec:geometry}.
\end{proof}

Thus a \(t\)-design is a cubature formula on the projective space of rotations. It is exact precisely for responses of spin at most \(t\). For example, the ideal two-use mean \(t_2\) and the dephasing discrepancy \(r_\lambda\) of Section~\ref{sec:q-coefficient} have spin at most two, so every \(2\)-design averages them exactly.

The bound for a general response is obtained by Favard's argument with one change. The optimal operator of Theorem~\ref{thm:even-sharp} replaces the projective Green kernel by a polynomial in its first variable. A design cannot see this polynomial part: averaged over the design, it becomes a constant. Only the residual of the kernel remains, and its integral norm is the sharp constant. For a weighted set put
\begin{equation}\label{eq:design-potential}
 \Phi_Q(V)=\sum_{i=1}^Nw_iK\bigl((\tfrac12\operatorname{Tr}(U_i^*V))^2\bigr),
 \qquad
 e(Q)=\min_{c\in\R}\int_{\SU}|\Phi_Q-c|\,d\sigma .
\end{equation}
Here \(K\) is the even Green kernel~\eqref{eq:even-kernel}, so \(\Phi_Q\) is the potential of unit charges placed at the gates of the set.

\begin{theorem}\label{thm:design-error}
Let \(Q\) be a weighted unitary \(t\)-design in \(\SU\). For every continuous even function \(f\) with \(\norm{\mathcal Lf}_\infty\le1\),
\begin{equation}\label{eq:design-error}
 \Bigl|Qf-\int_{\SU}f\,d\sigma\Bigr|
 \le e(Q)
 \le h_{t+1} .
\end{equation}
The first inequality is an equality for some such \(f\), and smooth functions in the class come arbitrarily close to it. For \(t=0\) the second inequality is an equality for every single gate.
\end{theorem}

\begin{proof}
By~\eqref{eq:even-potential}, \(f(U_i)-\bar f=\int K((U_i\cdot V)^2)\mathcal Lf(V)\,d\sigma(V)\), where \(U\cdot V=\tfrac12\operatorname{Tr}(U^*V)\). Averaging over the design gives
\begin{equation}\label{eq:design-representation}
 Qf-\bar f=\int_{\SU}\Phi_Q\,\mathcal Lf\,d\sigma
          =\int_{\SU}(\Phi_Q-c)\,\mathcal Lf\,d\sigma
\end{equation}
for every constant \(c\), since \(\mathcal Lf\) has mean zero. This proves the first inequality in~\eqref{eq:design-error}.

For the second, take the interpolating polynomial \(p_t\) of~\eqref{eq:even-error} with \(k=t+1\). For fixed \(V\) the function \(U\mapsto p_t((U\cdot V)^2)\) is an even polynomial of degree at most \(2t\). By Lemma~\ref{lem:design-cubature} its average over the design equals its Haar average \(c_0=\int_0^1p_t\,d\nu\), which does not depend on \(V\). Hence
\[
 \Phi_Q(V)-c_0=\sum_{i=1}^Nw_i\,(K-p_t)\bigl((U_i\cdot V)^2\bigr).
\]
By rotation invariance each term has integral norm \(h_{t+1}\), and the weights have sum one. Thus \(e(Q)\le h_{t+1}\).

The potential \(\Phi_Q\) is real-analytic away from the gates and unbounded near them, so each of its level sets has measure zero. Let \(c_*\) be its median and put \(g=\sgn(\Phi_Q-c_*)\). Then \(g\) is even, bounded by one and has mean zero. Its potential \(f(U)=\int K((U\cdot V)^2)g(V)\,d\sigma(V)\) is continuous with \(\mathcal Lf=g\), and~\eqref{eq:design-representation} gives \(Qf-\bar f=\int|\Phi_Q-c_*|\,d\sigma=e(Q)\), since the median minimizes the integral in~\eqref{eq:design-potential}. Smoothing by rotations as in the proof of Theorem~\ref{thm:even-sharp} gives smooth functions with errors arbitrarily close to \(e(Q)\). For \(t=0\) and a single gate, \(\Phi_Q\) is a rotated copy of \(K\), and \(e(Q)\) is the distance of \(K\) from the constants in \(L^1(\nu)\), which is \(h_1\).
\end{proof}
By homogeneity, for an arbitrary continuous even \(f\) with bounded Laplacian the right sides are multiplied by \(\norm{\mathcal Lf}_\infty\).

For Sobolev classes on spheres, the worst-case error of equal-weight cubature and the role of spherical designs were studied by Brauchart, Saff, Sloan and Womersley~\cite{BrauchartSaffSloanWomersley}. Here the class is defined by a bounded Laplacian in the uniform norm, and the sharp constant comes from the Green kernel. The quantity \(e(Q)\) is the deviation of the potential of unit charges at the gates; discrete energies and potentials of such configurations are the subject of the monograph of Borodachov, Hardin and Saff~\cite{BorodachovHardinSaff}.

The bound \(h_{t+1}\) holds for all \(t\)-designs at once and is the sharp error of the optimal linear method at the same cutoff. It is half of the bound \(2h_{t+1}\) obtained by applying the design to \(f-T_{t+1}f\) and to its Haar mean separately. The number \(e(Q)\) belongs to a particular set of gates. For the rotation groups below, Monte Carlo integration gives values of \(e(G)\) about one quarter smaller than \(h_{t+1}\); in what follows we use only the certified bound \(h_{t+1}\). With signed weights the same proof gives the factor \(\sum_i|w_i|\) in front of \(h_{t+1}\).

Responses of controlled gates, such as the Hadamard test of Section~\ref{sec:geometry}, are not even. Split such a response as \(f=f_++f_-\) with \(f_\pm(U)=(f(U)\pm f(-U))/2\). If the lifts of the gates to \(\SU\) are closed under \(U\mapsto-U\), as the binary groups behind Table~\ref{tab:polyhedral-designs} are (Appendix~\ref{app:computation}), its average of the odd part \(f_-\) vanishes, and so does the Haar mean of \(f_-\). Theorem~\ref{thm:design-error} then applies to \(f_+\) without change, with \(\norm{\mathcal Lf_+}_\infty\le\norm{\mathcal Lf}_\infty\).

The size of a design and its accuracy are related in the expected way. If \(s=\lfloor t/2\rfloor\) and \(N<\dim\mathcal V_{s+1}=(s+1)(2s+1)(2s+3)/3\), some nonzero \(p\in\mathcal V_{s+1}\) vanishes at all gates. Then \(|p|^2\in\mathcal V_{t+1}\) has average zero over the set and a positive Haar mean, so the set is not a \(t\)-design. Conversely, a spherical design of strength \(2t\) on \(S^3\) is exact for all polynomials of degree at most \(2t\), and in particular is a unitary \(t\)-design. Bondarenko, Radchenko and Viazovska proved that spherical designs of strength \(2t\) on \(S^3\) exist with every number of points \(N\ge Ct^3\)~\cite{BondarenkoRadchenkoViazovska}. By~\eqref{eq:spin-error-bound} their error is at most \(\beta_3/(4(t+1)^2)\), which is of order \(N^{-2/3}\) in the number of gates. No rule using \(N\) values does better in order. Indeed, take \(k=2N\) in the proof of Theorem~\ref{thm:q-rates} and use the even bumps \(u_i\) of that proof before they are paired by \(J\). Their supports are disjoint, so at most \(N\) of them are nonzero at some gate. The sum of the remaining ones, divided by a fixed constant, is a response in the class that vanishes at every gate, as does the zero response. Its mean is a fixed multiple of \(Nh^5\), that is, of \(N^{-2/3}\).

The designs used in practice are groups. If \(G\) is a finite subgroup of \(\SU\) and \(Q\) the uniform average over it, then \(Q\) applied to the matrix coefficients of the spin-\(j\) representation \(\pi_j\) gives the orthogonal projection onto the vectors fixed by \(G\). By Lemma~\ref{lem:design-cubature}, \(G\) is a \(t\)-design exactly when no spin \(j=1,\ldots,t\) has a nonzero fixed vector. The dimension of the fixed space is the average of the character~\eqref{eq:character-intro} over the group. A rotation through \(\phi\) has eigenvalues \(e^{\pm i\phi/2}\) in \(\SU\), so its spin-\(j\) character is \(\chi_{2j}(\phi/2)=\sin((2j+1)\phi/2)/\sin(\phi/2)\). For example, the single-qubit Clifford group consists of the \(24\) rotations of a cube. Apart from the identity it contains six quarter turns, nine half turns and eight rotations through \(2\pi/3\). The fixed space of spin \(j\) therefore has dimension
\[
 \tfrac1{24}\bigl[(2j+1)+6\chi_{2j}(\pi/4)+9\chi_{2j}(\pi/2)+8\chi_{2j}(\pi/3)\bigr].
\]
This vanishes for \(j=1,2,3\) and equals one for \(j=4\). The Clifford group is therefore a unitary \(3\)-design and not a \(4\)-design, in agreement with Webb~\cite{WebbClifford} and Zhu~\cite{ZhuClifford}. The same calculation for the other rotation groups gives Table~\ref{tab:polyhedral-designs}. The four Pauli gates form a \(1\)-design, the twelve rotations of a tetrahedron a \(2\)-design, and the sixty rotations of an icosahedron a \(5\)-design.

\begin{table}[ht]
\centering
\caption{Rotation groups as unitary designs. The column \(h_{t+1}\) is the guaranteed error in~\eqref{eq:design-error}. The last column is the actual error for the response \(f_0\) of~\eqref{eq:q-calibration}, which has \(\|\mathcal L f_0\|_\infty=1\).}\label{tab:polyhedral-designs}
\begin{tabular}{@{}lrrrr@{}}
\hline
Gates & Number & \(t\) & \(h_{t+1}\) & Error for \(f_0\)\\
\hline
Pauli & 4 & 1 & 0.08647 & 0.01692\\
Tetrahedral & 12 & 2 & 0.04673 & 0.00027\\
Clifford & 24 & 3 & 0.02939 & 0.00027\\
Icosahedral & 60 & 5 & 0.01477 & \(2\cdot10^{-7}\)\\
\hline
\end{tabular}
\end{table}

For a mean sign response with \(\|\mathcal L f\|_\infty\le S\), Theorem~\ref{thm:design-error} and the certified value of \(h_4\) (Appendix~\ref{app:computation}) give the Clifford average an error of at most \(h_4 S\le 0.02939\,S\). The corresponding bound for the positive-outcome probability is \(0.014695\,S\). On the specific response \(f_0\), the error is only \(0.00027\) to the precision shown. For an extremal response, \(\mathcal L f\) changes sign where the potential \(\Phi_G\) crosses its median, as in the proof of Theorem~\ref{thm:design-error}.

The guarantee concerns the response as a function of the command. It therefore applies to a device whose response varies smoothly with the commanded rotation. As explained in Section~\ref{sec:q-coefficient}, a compilation rule can make the response discontinuous along a chart boundary. Then \(\norm{\mathcal Lf}_\infty\) is infinite, and a finite set of commands can miss the error entirely.

On \(n\) qubits the procedures cited above mostly use local designs: each qubit receives its own gate, drawn independently from a single-qubit design. Classical shadows with random Pauli measurements use independent single-qubit Clifford gates~\cite{HuangKuengPreskill}, and randomized compiling uses independent Pauli gates~\cite{WallmanEmerson}. A response is then a function \(f(U_1,\ldots,U_n)\) on the product of \(n\) copies of \(\SU\). We assume it is even in each variable and write \(\mathcal L_i=-\Delta_i\) for the negative Laplacian in the \(i\)th variable.

The errors of the separate qubits simply add up: each qubit contributes the one-qubit error of Theorem~\ref{thm:design-error}, multiplied by the bound on the Laplacian in its own variable. Indeed, let \(Q_i\) be a weighted unitary \(t_i\)-design for the \(i\)th qubit and \(Q=Q_1\otimes\cdots\otimes Q_n\). Write \(A_i\) for the average over \(Q_i\) and \(H_i\) for the Haar average, each acting in the \(i\)th variable. These operators commute, and
\[
 Qf-\int f
  =\sum_{i=1}^n A_1\cdots A_{i-1}(A_i-H_i)H_{i+1}\cdots H_nf .
\]
In the \(i\)th term the other operators average the remaining variables against probability measures; they commute with \(\mathcal L_i\) and do not increase the uniform norm. The \(i\)th term is therefore the error of \(Q_i\) on an even function of one gate, and Theorem~\ref{thm:design-error} gives
\begin{equation}\label{eq:local-designs}
 \Bigl|Qf-\int f\,d\sigma^{\otimes n}\Bigr|
 \le\sum_{i=1}^n h_{t_i+1}\norm{\mathcal L_if}_\infty .
\end{equation}

The number of qubits enters only through the number of terms. A response that depends on \(n_0\) of the gates contributes only \(n_0\) terms, whatever the size of the processor. For example, a product of \(n_0\) single-qubit expectations \(\operatorname{Tr}(O_iU_i\rho_iU_i^*)\) with traceless \(O_i,\rho_i\) has \(\mathcal L_if=8f\) in each active variable, so the right side of~\eqref{eq:local-designs} grows linearly in~\(n_0\).

Global designs behave differently. The \(n\)-qubit Clifford group is a unitary \(3\)-design on \(\mathrm{SU}(2^n)\)~\cite{WebbClifford,ZhuClifford}. The argument of Theorem~\ref{thm:design-error} applies on any compact group. Use the bi-invariant metric \(\tfrac12\operatorname{Re}\operatorname{Tr}(X^*Y)\), which gives the metric of the unit sphere for \(N=2\). If \(f\) is invariant under the centre and \(p\) is a central function in the span of the characters of the irreducible components of \(U^{\otimes t}\otimes\overline U^{\otimes t}\), then a \(t\)-design averages \(p(U_i^{-1}V)\) to a constant. The error is therefore at most the integral distance of the centrally averaged Green kernel of \(\mathrm{SU}(N)\) from such functions, times \(\norm{\mathcal Lf}_\infty\). For \(N=2\) this distance is \(h_{t+1}\).

The normalization of the Laplacian also changes with \(N\). For traceless Hermitian \(O,\rho\), the response \(f(U)=\operatorname{Tr}(OU\rho U^*)\) on \(\mathrm{SU}(N)\) satisfies \(-\Delta f=4N\,f\). For \(N=2\) this is the eigenvalue \(8\) of~\eqref{eq:q-radial-t}. Thus, with \(N=2^n\), a bound on the global Laplacian of an \(n\)-qubit response is \(4N=2^{n+2}\) times its size even for the simplest observable. This is the reason why the local form~\eqref{eq:local-designs} is the useful one for engineering.

\section{Certifying a compressed circuit}\label{sec:compression}

Lossy compression of quantum circuits, as developed by Haiqu~\cite{Bohun,HaiquAcademic}, replaces a deep circuit by a shallower one with optimized parameters whose output is close to the original output. A circuit that will run many times is worth shortening in this way. Typically the circuit belongs to a family. A command \(U\) loads a data point, sets a variational layer, or applies a calibrated gate, while the rest of the circuit is fixed. The user of a compressor then needs to know how much the expected output can change, either at the worst command or on average over all of them. Both questions concern the difference of two responses on the group, and both can be settled on an emulator, where responses are computed exactly and without shot noise.

The original family has a structure that a compressor need not preserve. Suppose that a circuit on any number of qubits uses the single-qubit command \(U\) exactly \(m\) times, and that all other operations, including noise channels, do not depend on \(U\). Write \(f(U)\) for the expected value of a fixed observable at the output.

\begin{lemma}\label{lem:spin-count}
Then \(f\in\mathcal V_{m+1}\): the response has spin at most \(m\). If some uses of \(U\) are controlled by other qubits, \(f\) is still a polynomial of degree at most \(2m\) on \(S^3\), but it need not be even.
\end{lemma}

\begin{proof}
An uncontrolled use acts on the density matrix by \(\rho\mapsto(U\otimes I)\rho(U\otimes I)^*\). Each matrix entry of the result is a linear combination of products of one entry of \(U\) and one entry of \(\overline U\). The fixed operations are linear. Hence \(f\) is a linear combination of products of \(m\) entries of \(U\) and \(m\) entries of \(\overline U\), which by the proof of Lemma~\ref{lem:design-cubature} lie in \(\mathcal V_{m+1}\). A controlled use acts through the block matrix \(I\oplus U\), whose entries have degree zero or one. The product then has degree at most \(2m\), with odd terms allowed.
\end{proof}

The lemma has two practical consequences. First, an \(m\)-design averages the original response exactly. The two-use mean \(t_2\) of the introduction is the case \(m=2\). Second, the largest value of \(|f|\) over all commands can be certified from finitely many exact values. If \(X\) is a projective net with covering radius at most \(1/(4\mathrm em)\), then \(\norm{f}_\infty\le2\max_{x\in X}|f(x)|\) by~\eqref{eq:q-norming} with \(L=2m\). For controlled uses the same holds with a net on \(S^3\) itself.

A compressor that rebuilds the whole circuit for each command destroys this structure. The compressed response \(\tilde f\) need not be a polynomial in \(U\), and the discrepancy \(d=\tilde f-f\) is a general function on the group. We assume that \(d\) is even and that \(\norm{\mathcal Ld}_\infty\le S\). By~\eqref{eq:pulse-curvature} this is a bound on the curvature of the loss along the three rotation axes. On an emulator these curvatures can be sampled by finite differences at any command. Such sampling is a diagnostic, not a proof of the bound.

Under this assumption the loss can be certified from finitely many runs of the two circuits, in two ways. The first bounds the largest loss over all commands by its values on a finite net of commands. Let \(k\ge2\), \(L=2k-2\), and let \(X\) be a projective net with covering radius at most \(1/(2\mathrm eL)\). The polynomial \(v=T_kd\) satisfies \(\norm{d-v}_\infty\le Sh_k\) by Theorem~\ref{thm:even-sharp}, and it has degree at most \(L\), so~\eqref{eq:q-norming} gives \(\norm{v}_\infty\le2\max_X|v|\le2\max_X|d|+2Sh_k\). Hence
\begin{equation}\label{eq:compression-uniform}
 \norm{d}_\infty\le 2\max_{x\in X}|d(x)|+3S\,h_k .
\end{equation}
The second bounds the error of the mean loss computed over a design. By Theorem~\ref{thm:design-error}, for every unitary \(t\)-design \(Q\),
\begin{equation}\label{eq:compression-mean}
 \Bigl|Qd-\int_{\SU}d\,d\sigma\Bigr|\le h_{t+1}S .
\end{equation}

For a general smooth discrepancy the first term of~\eqref{eq:compression-uniform} uses values of both circuits on the net. The grid of Appendix~\ref{app:value} has at most \(4(10L+1)^3\) nodes, giving \(37\,044\) evaluations per circuit for \(k=2\) and \(275\,684\) for \(k=3\) before merging repeated nodes. The corresponding smoothness allowances are \(3Sh_2<0.26\,S\) and \(3Sh_3<0.15\,S\). For the signed mean discrepancy, evaluating both circuits at the \(24\) Clifford commands takes \(48\) runs and gives a certified error of at most \(h_4 S\le 0.02939\,S\) by~\eqref{eq:compression-mean}.

Suppose now that the compressed circuit retains the command as an uncontrolled gate, uses it \(\tilde m\) times and compresses only the fixed operations. If the \(m\) original uses are also uncontrolled, Lemma~\ref{lem:spin-count} gives \(d\in\mathcal V_{M+1}\) for \(M=\max\{m,\tilde m\}\). In this case a \(2M\)-design determines the whole discrepancy. Indeed, for any Haar-orthonormal basis \(\{\varphi_j\}\) of \(\mathcal V_{M+1}\), the product \(d\,\overline{\varphi_j}\) has spin at most \(2M\). Hence its coefficient is the finite sum
\begin{equation}\label{eq:design-reconstruct}
 c_j=\int d\,\overline{\varphi_j}\,d\sigma
     =\sum_i w_i d(U_i)\overline{\varphi_j(U_i)},\qquad
 d=\sum_j c_j\varphi_j .
\end{equation}
The same design integrates \(d^2\) exactly. Once the coefficients are known, one evaluates the polynomial on a net \(X\) of radius at most \(1/(4\mathrm e M)\) for \(M\ge1\). The bound \(\|d\|_\infty\le2\max_X|d|\) follows from~\eqref{eq:q-norming} with no further circuit evaluations. The twelve tetrahedral commands suffice for \(M=1\), and the sixty icosahedral commands suffice for \(M=2\).

We tested this rule on a small Hamiltonian simulation. Take four qubits in a line and the Ising Hamiltonian \(H=\sum_{i=1}^3Z_iZ_{i+1}+\sum_{i=1}^4X_i\). The original circuit applies the command \(U\) to the first qubit, then forty first-order Trotter steps for \(e^{-\mathrm iH}\), then \(U\) again, and measures \(Z\) on the first qubit. Each Trotter step contains three two-qubit \(ZZ\) rotations, so the circuit has \(240\) CNOT gates. The compressed circuit keeps both uses of \(U\) and replaces the forty steps by three layers of the same shape. Their \(21\) angles were fitted to the forty steps, which leaves a normalized trace infidelity of \(1.2\cdot10^{-3}\) and \(18\) CNOT gates. Thus \(m=\tilde m=M=2\). A least-squares fit by polynomials of degree four on \(S^3\) reproduces both responses to within \(10^{-15}\), as Lemma~\ref{lem:spin-count} predicts. The original response has largest absolute value \(0.97\). Table~\ref{tab:compression} lists the certificates for the loss \(d\).

\begin{table}[ht]
\centering
\caption{Certifying a \(13\)-fold compression of a four-qubit Trotter circuit. The Haar moments~\eqref{eq:exp-map-haar-moment} integrate the polynomial losses exactly. Numerical entries, including the value of the mesh bound, are rounded.}\label{tab:compression}
\begin{tabular}{@{}lr@{}}
\hline
Quantity & Value\\
\hline
largest \(|d|\) on the \(275\,684\)-point grid & \(0.012500\)\\
uniform upper bound \(2\max_X|d|\) & \(0.025000\)\\
largest \(|d|\) found by local maximization & \(0.012502\)\\
\hline
Haar mean of \(d\), exact & \(-7.108\cdot10^{-5}\)\\
\quad Pauli average (a \(1\)-design) & \(-2.132\cdot10^{-4}\)\\
\quad tetrahedral, Clifford and icosahedral averages & \(-7.108\cdot10^{-5}\)\\
\hline
Haar mean of \(d^2\), exact & \(2.5382\cdot10^{-5}\)\\
\quad Clifford average (a \(3\)-design) & \(2.6677\cdot10^{-5}\)\\
\quad icosahedral average (a \(5\)-design) & \(2.5382\cdot10^{-5}\)\\
\hline
\end{tabular}
\end{table}

Here \(\dim\mathcal V_3=1+9+25=35\). The \(60\) icosahedral commands thus recover all \(35\) coefficients of \(d\) by~\eqref{eq:design-reconstruct} from \(120\) exact emulator values, one per circuit at each command. These values suffice both for the mean-square loss and for the uniform bound in Table~\ref{tab:compression}. All \(275\,684\) grid evaluations in the latter calculation are evaluations of the reconstructed quartic. The same count of circuit evaluations applies when the command is embedded in a larger register. For the signed loss, every \(2\)-design is exact. For its square, the Clifford average exceeds the Haar value by about five percent, while the icosahedral average is exact.

The four-qubit example was computed on an ordinary computer. Since the certificate requires only \(120\) exact evaluations of the two circuits, whatever the number of qubits, it would be interesting to run the same procedure for much larger circuits on a state-vector emulator, for example on the emulator of Quantum Lab Azores, which runs mpiQulacs on eight Fujitsu A64FX nodes for up to \(33\) qubits.

A compressor optimizes a distance between circuits, while the user needs the loss in an observable. The standard bridge between the two is crude. When the command is kept as a gate, the output states of the two circuits differ by \((U\otimes I)(\widetilde W-e^{\mathrm i\alpha}W)(U\otimes I)|0\rangle\) for every phase \(\alpha\). Since \(|\langle\tilde\psi|O|\tilde\psi\rangle-\langle\psi|O|\psi\rangle|\le2\norm{O}\,\norm{\tilde\psi-\psi}\) for unit vectors, every command satisfies
\begin{equation}\label{eq:operator-loss}
 |d(U)|\le2\norm{O}\min_\alpha\norm{\widetilde W-e^{\mathrm i\alpha}W}.
\end{equation}
In the example the operator distance is \(0.0882\), so~\eqref{eq:operator-loss} gives \(0.176\). The uniform bound from the measured observable is approximately \(0.025\), about seven times smaller. Local maximization found a loss of \(0.012502\).

The emulator certifies the loss before the compressed circuit leaves the laboratory. A device can confirm the certificate only within the statistics of its readouts, and the numbers are instructive. Consider first a single command. If each circuit is run \(m\) times and returns signs \(\pm1\), the difference of the two sample means has variance at most \(2/m\). To make three standard errors at most a loss of size \(\delta\), it suffices to take \(m\ge18/\delta^2\) runs of each circuit. For a loss of \(\delta=0.0125\) of the example this is \(115\,200\) runs, which is feasible.

The mean-square loss can be estimated from paired readouts. At a command \(U\), take two independent signs \(\tilde X,\tilde X'\) from the compressed circuit and two independent signs \(X,X'\) from the original one. Then
\begin{equation}\label{eq:paired-compression}
 \mathbb E\bigl[(\tilde X-X)(\tilde X'-X')\mid U\bigr]=d(U)^2 ,
\end{equation}
since the two factors are independent with mean \(d(U)\) each. When \(M\le2\), the icosahedral design averages \(d^2\) exactly, so the mean of these products over the sixty icosahedral commands is an unbiased estimate of \(\int d^2\,d\sigma\). Each product lies in \([-4,4]\). In the example its standard deviation is \(1.76\), while the quantity to be measured is \(2.5\cdot10^{-5}\). To estimate it within its own size at two standard errors takes about \(1.9\cdot10^{10}\) pairs, that is \(8\cdot10^{10}\) circuit runs, and Hoeffding's inequality for the same error with confidence \(0.95\) asks for ten times more.

With exact emulator values the finite design determines the loss polynomial and certifies a uniform bound. On a device the observations at those same commands are random. The budget above describes the precision of the estimator formed from individual readout pairs in~\eqref{eq:paired-compression}.

\section{Single-state responses on the Bloch sphere}\label{sec:hopf}

Many responses depend on a gate only through the state it prepares. This is the case for state preparation and data loading, where a command \(U\) produces the state \(U|0\rangle\) and everything afterwards is fixed, and for the single-use readout \(t(U)\) of Section~\ref{sec:q-coefficient}. Such responses are functions on the Bloch sphere, and for them the sharp constants are smaller and have a closed form.

After one ideal gate the prepared state determines its Bloch vector. After two applications, the response can vary along the fibre of gates giving that same first output, as the example in the introduction shows. The dependence on that one state gives a simple relation between the degrees and Laplacians on \(S^2\) and \(S^3\). We use both spheres with radius one.

For the first column \((z,w)\) in \eqref{eq:matrix} the Hopf map~\cite{Hopf1931} is
\begin{equation}\label{eq:hopf-map}
 h(z,w)=\big(2\operatorname{Re}(\bar z w),\,2\operatorname{Im}(\bar z w),
                           \,|z|^2-|w|^2\big).
\end{equation}
Its norm is one, and every point of \(S^2\) occurs. For example take \(|z|^2=(1+h_3)/2\) and choose the relative phase to obtain \(h_1+ih_2\). Each fibre consists of the vectors \(e^{is}(z,w)\). This is the Bloch vector of the output pure state~\cite[\S\,1.2]{NielsenChuang}; see also~\cite[\S\,3.1]{Schmocker} for this representation in the context of quantum input spaces. Every pullback \(F\circ h\) is even. Many even functions depend on the gate's action on other states as well.

To use the sharp constants of the two-sphere we need two facts, collected in the next lemma: which polynomials on the gates depend only on the prepared state, and how the Laplacian changes when a function on the Bloch sphere is regarded as a function of the gate.

\begin{lemma}\label{lem:hopf}
The functions in \(\mathcal V_k\) that are constant along the fibres are exactly \(q\circ h\), where \(q\) is a spherical polynomial on \(S^2\) of degree at most \(k-1\).
With the Laplacians of the unit spheres, for \(C^2\) inputs,
\begin{equation}\label{eq:hopf-lap}
  \Delta_{S^3}(F\circ h)=4(\Delta_{S^2}F)\circ h .
\end{equation}
The same formula holds for continuous \(F\) with bounded Laplacian.
\end{lemma}
\begin{proof}
Each \(q\circ h\) belongs to \(\mathcal V_k\) since the components of \(h\) are quadratic. Conversely, let \(p\in\mathcal V_k\) be constant on the fibres. Averaging \(p\) under \((z,w)\mapsto e^{is}(z,w)\) leaves \(p\) unchanged and retains only monomials with equal total degrees in the variables and their conjugates. Each such monomial is a product of at most \(k-1\) factors from \(|z|^2,|w|^2,\bar z w,z\bar w\). All four factors are affine functions of \(h\).

For the differential formula write the three real quadratic forms as \(h_i(x)=x^\top A_i x\) in the real coordinates of \((z,w)\). Direct multiplication of the displayed forms gives symmetric traceless matrices with \(A_i^2=I\) and \(A_i A_j+A_j A_i=0\) for \(i\ne j\). On \(S^3\) it follows that
\begin{equation}\label{eq:hopf-coordinate-identities}
   \Delta h_i=-8h_i,\qquad
   \nabla h_i\cdot\nabla h_j
     =4(\delta_{ij}-h_i h_j).
\end{equation}
The second identity follows by projecting \(2A_i x\) onto the tangent space at \(x\). Use the \(C^2\) extension \(v\mapsto F(v/|v|)\) near \(S^2\) in the chain rule. At \(v\in S^2\) its spherical Laplacian is
\(\sum_{ij}(\delta_{ij}-v_i v_j)\partial_{ij}F
      -2\sum_i v_i\partial_i F\).
Substitution proves \eqref{eq:hopf-lap}. For general continuous \(F\) with bounded Laplacian, smooth \(F\) by convolution with approximate identities on \(\SO\). The smoothed functions converge uniformly, their Laplacians converge weak-* in \(L^\infty(S^2)\), and their pullbacks have the corresponding weak-* convergence on \(S^3\). For the last assertion integrate a test function first over the phase fibre; the pushforward of normalized Haar is the rotation-invariant probability on \(S^2\). The identity follows by testing and passing to the limit.
\end{proof}

In particular, if \(Y_j\) is a spherical harmonic of degree \(j\) on the unit two-sphere, the eigenvalue of \(Y_j\circ h\) for \(-\Delta_{S^3}\) is \(4j(j+1)\). Thus \(Y_j\circ h\) has spherical degree \(2j\) on \(S^3\) and spin \(j\). The third coordinate of \(h\) gives the spin-one observable used below.

For functions on the two-sphere our earlier work gives the answer in closed form. Write \(H(n,S^2)\) for the sharp constant of the Bohr problem on \(S^2\): the supremum of \(\norm{F}_\infty\) over \(F\) with \(\norm{\Delta_{S^2}F}_\infty\le1\) and no spherical harmonics of degree below \(n\). By~\cite{SolyanikLaplacian},
\begin{equation}\label{eq:S2-bohr}
 H(n,S^2)=2\sum_{j=0}^\infty\frac{1}{(2j+1)\bigl((n+1)^2(2j+1)^2-1\bigr)},
 \qquad
 \lim_{n\to\infty}n^2H(n,S^2)=\frac74\zeta(3),
\end{equation}
and by~\cite{SolyanikLaplacian} the same number is the smallest worst-case error of a linear method with values in polynomials of degree below \(n\). The Hopf map carries this answer to the gates. Consider the responses that depend on \(U\) only through the prepared state \(U|0\rangle\), that is, \(f=F\circ h\) with \(\norm{\mathcal Lf}_\infty\le1\). Let \(h_k^{\rm st}\) be the smallest worst-case error on these responses of a linear method with values in \(\mathcal V_k\) that reproduces constants. We show that
\begin{equation}\label{eq:state-sharp}
 h_k^{\rm st}=\tfrac14H(k,S^2);
\end{equation}
in particular \(h_1^{\rm st}=\tfrac14\log2\) and \(k^2h_k^{\rm st}\to\tfrac{7}{16}\zeta(3)\), by summing the series in~\eqref{eq:S2-bohr} and by its limit. By~\eqref{eq:hopf-lap}, \(\norm{\mathcal Lf}_\infty=4\norm{\Delta_{S^2}F}_\infty\). Applying the optimal method of~\cite{SolyanikLaplacian} to \(F\) and composing with \(h\) gives an output \(q\circ h\) with \(\deg q<k\), which lies in \(\mathcal V_k\) by Lemma~\ref{lem:hopf}, with error at most \(\tfrac14H(k,S^2)\). Conversely, averaging the output of any admissible method over the phases \((z,w)\mapsto e^{is}(z,w)\) does not increase its error on \(F\circ h\), and by Lemma~\ref{lem:hopf} turns it into a linear method on \(S^2\) with values in polynomials of degree below \(k\). By~\cite{SolyanikLaplacian} its error is at least \(\tfrac14H(k,S^2)\).

\begin{table}[ht]
\centering
\caption{Sharp linear errors for responses of a prepared state and for general responses, at the same spin cutoff.}\label{tab:state-constants}
\begin{tabular}{@{}rrr@{}}
\hline
\(k\) & \(h_k^{\rm st}=\tfrac14H(k,S^2)\) & \(h_k\)\\
\hline
1 & 0.17329 & 0.22154\\
2 & 0.06541 & 0.08647\\
3 & 0.03496 & 0.04673\\
4 & 0.02187 & 0.02939\\
5 & 0.01501 & 0.02022\\
\hline
\end{tabular}
\end{table}

A response that depends only on the prepared state thus enjoys a guarantee about one quarter smaller than that of a general response with the same bound on the Laplacian, and the guarantee is given by an explicit series. The optimal method is also explicit. On \(S^2\) the best approximant of the Green kernel comes from trigonometric interpolation at the equally spaced angles \(j\pi/(k+1)\), \(j=1,\ldots,k\)~\cite{SolyanikLaplacian}, so no moment equations have to be solved. For a procedure that loads a function into a prepared state, the same reasoning as in~\eqref{eq:parameter-budget} gives a budget of \(k^2\) coefficients on the sphere, with \(k\approx(7\zeta(3)S/(16\varepsilon))^{1/2}\).

\section{The readout error of a two-gate circuit}\label{sec:q-coefficient}

In this section we work out the simplest circuit, the one of the introduction: a qubit is prepared in \(|0\rangle\), the gate \(U\) is applied twice, and \(Z\) is measured. We write its ideal readout, explain what continuity and a bounded Laplacian of the response mean for a real compiled device, and compute the error \(r\) and the bound on its Laplacian exactly for a dephasing channel between the two gates. At the end of the section we introduce the readout after a single use of the gate and its coefficient \(a(f)\), the quantity studied in Section~\ref{sec:q-optimal}.

Fix \(|0\rangle\) as input and the Pauli observable \(Z=\operatorname{diag}(1,-1)\) as readout. Let \(R=R(U)\) be the rotation of the Bloch vector induced by \(U\), with indices \(1,2,3\) corresponding to \(x,y,z\). An ideal circuit applies \(U\) twice with no intervening measurement. Its two-use and one-use means are respectively
\begin{equation}\label{eq:two-use-signal}
 t_2(U)=e_z^{\mathsf T}R^2e_z=2|(U^2)_{11}|^2-1,
 \qquad t(U)=e_z^{\mathsf T}Re_z=R_{33}.
\end{equation}
For the pair of commands in the introduction \(t\) vanishes at both gates while \(t_2\) takes opposite extreme values. By Lemma~\ref{lem:spin-count} with \(m=2\), \(t_2\) has spin at most two. Spin two occurs already on rotations about the \(y\)-axis, where \(t_2(R_y(\theta))=\cos(2\theta)\). Keeping this known part of the signal exactly leaves the available spin cutoff for learning the discrepancy \(r=f-t_2\).

For each sampled command the processor prepares the state afresh, executes the specified gate twice and returns one sign. Any modeled interval between the gates belongs to the same circuit. The smoothness assumption in~\eqref{eq:shot-map-uniform} concerns the stationary response for this particular compilation and sequence. If a probability for a new command is required, the raw prediction is \((1+\widehat f_k(U))/2\), which may be clipped to \([0,1]\) for reporting. If a tolerance must hold for every command, \eqref{eq:shot-map-discrepancy} instead bounds the largest deviation of the device from its ideal readout. The preparation and the measured observable remain fixed in both specifications.

The continuity at issue is continuity in the commanded rotation. For example, in a \(z\)-\(y\)-\(z\) Euler compilation, two angle triples at middle angle zero can specify the same rotation through their common sum of outer angles. Physical errors in distinct pulse recipes may give different limiting readouts as the command approaches that rotation. If the limiting means differ by \(\delta>0\), every continuous predicted mean has uniform error at least \(\delta/2\), since it has the same limit along both paths. At an isolated command a changed readout can also escape Haar sampling. The bound \(S_r\) therefore requires independent information about the compiled response, as in the channel calculation below. On hardware one can compare readings approaching a chart boundary from both sides and report the held-out mean-square error on the actual distribution of commands. Agreement on a finite set by itself supplies no upper bound for \(S_r\).

Consider a dephasing channel between the two gates. It maps the Bloch vector \((x,y,z)\) to \((\lambda x,\lambda y,z)\), with \(0\le\lambda\le1\). It can be realized by applying \(Z\) with probability \((1-\lambda)/2\) and the identity otherwise. For this model the mean sign and its discrepancy are
\begin{equation}\label{eq:dephasing-model}
 f_\lambda=e_z^{\mathsf T}R\,\operatorname{diag}(\lambda,\lambda,1)R e_z
           =\lambda t_2+(1-\lambda)t^2,\qquad
 r_\lambda=(1-\lambda)(t^2-t_2).
\end{equation}
On the two test gates of the introduction the measured means are \(-\lambda\) and \(+\lambda\).

For this channel the largest probability discrepancy and the exact smoothness bound are
\begin{equation}\label{eq:dephasing-smoothness}
 \mathcal P_\lambda=\frac12\|r_\lambda\|_\infty=\frac{1-\lambda}{2},
 \qquad \|\mathcal L r_\lambda\|_\infty=\frac{50}{3}(1-\lambda).
\end{equation}
To see this, use \(U=U(z,w)\) from \eqref{eq:matrix} and put \(\zeta=2\operatorname{Re}(z^2)\). Multiplying \(U\) by itself gives \((U^2)_{11}=z^2-|w|^2\), hence \(t^2-t_2=(1-t)\zeta\). Since \(t=2|z|^2-1\), one has \(|\zeta|\le1+t\), with equality at each \(t\) for real \(z\). Thus the largest modulus of \((1-t)\zeta\) is one. Both \(\zeta\) and \(t\) are degree-two harmonics and have Laplacian \(-8\) times the function. Projecting their Euclidean gradients onto the tangent space gives \(\nabla t\cdot\nabla\zeta=4(1-t)\zeta\). The product rule now gives \(\Delta[(1-t)\zeta]=8(3t-2)\zeta\). The maximum of \(|(3t-2)(1+t)|\) on \([-1,1]\) is \(25/12\), attained at \(t=-1/6\). This proves \eqref{eq:dephasing-smoothness}. The discrepancy in this model itself has spins at most two. The Favard operator minimizes worst-case uniform error over the smoothness class and need not reproduce each retained spin.

For comparison with the two-gate map, consider a single application of the gate and a single scalar summary of its readout.
Write $e$ for the identity matrix. Prepare $|0\rangle$, apply $U$ and measure $Z$, as in~\eqref{eq:two-use-signal}. The outcomes $1$ and $-1$ have probabilities $|U_{11}|^2$ and $1-|U_{11}|^2$, so the one-use mean is
\begin{equation}\label{eq:q-t}
 t(U)=2|U_{11}|^2-1=\tfrac12\operatorname{Tr}(ZUZU^*).
\end{equation}
Conjugation acts by rotations on the real space of traceless Hermitian
$2$-by-$2$ matrices, with inner product $\operatorname{Tr}(AB)/2$.
Consequently $t$ is a diagonal matrix coefficient of the spin-one
representation. For a real response $f$ we study the scalar
\begin{equation}\label{eq:q-functional}
 a(f)=\int_{\mathrm{SU}(2)} f(U)t(U)\,d\sigma(U).
\end{equation}
We use the matrix coefficient as given in \eqref{eq:q-t}, with no
dimension factor. 

Two elementary identities specify the normalizations. Identify the first
column of $U$ with $(z_1,z_2)\in S^3\subset\mathbb C^2$. Put
$z_1=\sqrt u\,e^{i\alpha}$ and $z_2=\sqrt{1-u}\,e^{i\beta}$.
For $0<u<1$ the diagonal entries of the metric of the unit sphere in
$(u,\alpha,\beta)$ are $1/(4u(1-u)),u,1-u$. Their product is $1/4$.
Haar probability is consequently
$du\,d\alpha\,d\beta/(2\pi)^2$, where $0<u<1$ and both angles have
period $2\pi$. It follows that $t=2u-1$ is uniform on $[-1,1]$, with
$\int t\,d\sigma=0$ and $\int t^2\,d\sigma=1/3$.
These moments give the coefficient a direct meaning for a gate calibration. Fit a straight line \(b+C t(U)\) to the mean response \(f(U)\), weighting rotations by Haar probability. The intercept is the Haar mean of \(f\), and the slope is
\begin{equation}\label{eq:readout-gain}
              C=\frac{\int_{\SU} f(U)t(U)\,d\sigma(U)}
                      {\int_{\SU}t(U)^2\,d\sigma(U)}=3a(f).
\end{equation}
For an ideal readout \(C=1\). The same interpretation applies to a different fixed pair of axes. If \(\mathbf n\) is the preparation direction, \(\mathbf m\) the readout direction and \(R(U)\) the rotation induced by conjugation on Pauli matrices, the ideal mean is \(\mathbf m^{\mathsf T}R(U)\mathbf n\). Fixed rotations of the two axes reduce this expression to \(t\) by left and right translation. We now show that the approximation bound does not change under such a change of axes. If fixed gates \(A\) and \(B\) are placed before and after the command, the circuit executes \(AUB\) instead of \(U\), and the operator of Theorem~\ref{thm:even-sharp} follows this change: for \(f_{A,B}(U)=f(AUB)\),
\begin{equation}\label{eq:frame-covariance}
 T_k(f_{A,B})(U)=(T_k f)(AUB).
\end{equation}
Indeed, the kernel in~\eqref{eq:even-method} depends only on \(\tfrac12\operatorname{Tr}(U^*V)\), which does not change under \((U,V)\mapsto(AUB,AVB)\), and the Laplacian and Haar measure are invariant under both translations. Hence the bound of Theorem~\ref{thm:even-sharp} is the same for every pair of axes. The spin \(1/2\) labels the qubit state, while spin one labels the rotation of the measured Pauli observable.

For a twice differentiable function $F$ of one real variable the metric
also gives
\begin{equation}\label{eq:q-radial-t}
 \Delta(F(t))=4(1-t^2)F''(t)-8tF'(t).
\end{equation}
In particular, for $f_0(U)=\exp(t(U))/(8\mathrm e)$, we have
\begin{equation}\label{eq:q-calibration}
 \|\mathcal L f_0\|_\infty=1,\qquad a(f_0)=\frac1{8\mathrm e^2}.
\end{equation}
Indeed, the multiplier of $F$ on the right of \eqref{eq:q-radial-t}
for $F(t)=\exp(t)/(8\mathrm e)$ is $8-4(t+1)^2$.
Its absolute value is at most $8$ and attains $8$ at $t=1$.
The coefficient follows by integrating $t\exp(t)$ on $[-1,1]$.
Subtracting $f_0(e)$ preserves both assertions. This one-variable example
provides a reference value for testing an implementation.

\section{The complexity of a coefficient with numerical value access}\label{sec:q-optimal}

The two-use experiment asks for the response at each command. Its worst probability discrepancy in the dephasing model is \((1-\lambda)/2\) by~\eqref{eq:dephasing-smoothness}, although \(a(f_\lambda)=0\) for every \(\lambda\) (Appendix~\ref{app:value}). We now consider the separate task of estimating \(a(f)\), which gives the one-use gain \(3a(f)\) in~\eqref{eq:readout-gain}. A classical query returns a rounded numerical value of \(f\) at a selected rotation. A coherent query evaluates these values in a superposition of addresses and can be undone. A reversible model or a supplied value table can furnish this access. Building the circuit or populating the table has a separate cost.

We now regard $f$ as unknown. It is real-valued and even, is continuous
on the unit sphere $S^3$, and has bounded Laplacian.
Our unit class is
\begin{equation}\label{eq:anchored-class}
 \mathcal F_*=\{f\text{ as above}: f(e)=0,\ \|\mathcal L f\|_\infty\le1\}.
\end{equation}
Anchoring selects a representative modulo additive constants. Since \(t\)
has mean zero, all representatives give the same \(a(f)\).
At spin cutoff one, \(T_1 f=\bar f\) and
\(\|f-\bar f\|_\infty\le h_1\) by \eqref{eq:even-first}. Thus on the
anchored class the value register need only cover the interval
\([-1/2,1/2]\), since
\begin{equation}\label{eq:q-range}
 |f(x)|=|f(x)-f(e)|\le 2h_1<1/2 .
\end{equation}

A deterministic query returns a rounded value of $f(x)$ at a selected
point $x$. For a quantum query choose a finite list $x_j$ and an encoding
of the above bounded interval to absolute accuracy $\delta$. On the
address and value registers the oracle acts by
\begin{equation}\label{eq:q-oracle}
 |j,z\rangle\longmapsto |j,z\mathbin{\oplus}\operatorname{enc}_\delta(f(x_j))\rangle .
\end{equation}
Both the list and the precision may depend on the requested error.
We count oracle calls, including those used to undo a computation.
Arithmetic on known data, preparing a known distribution and the size of
the registers are accounted for separately from query complexity.
Equation \eqref{eq:q-t} identifies the physical meaning of the weight.
The oracle in \eqref{eq:q-oracle} supplies the values of the unknown
response. If these values themselves have to be inferred from repeated
measurements, the cost of constructing that oracle is a separate question.

For each integer \(L\ge2\) put \(k=\lfloor L/2\rfloor+1\).
The polynomial \(T_k f\) of \eqref{eq:even-method} has degree at
most \(L\), is real for real inputs, and satisfies
\(\|f-T_k f\|_\infty\le h_k\). The sampling algorithm uses any known
upper bound \(b_L\) for \(h_k\). The certified upper endpoint in
Appendix~\ref{app:computation} is one choice. The bounds
\(h_k\le\beta_3/(4k^2)\) and \(\beta_3<3\) give the simpler
choice \(b_L=3/(4k^2)\) without computing the moment nodes. For
the order estimates we can also use
\begin{equation}\label{eq:q-classicalbound}
 b_L=B(L+1)^{-2},\qquad B=3 ,
\end{equation}
since \(2k\ge L+1\).

The coefficients of \(T_k f\) depend on moments of \(f\). With only
point values available, we construct a polynomial on a sampling net.
The uniform error bound for \(T_k f\) supplies the tolerance at the
nodes, and the spacing of the net controls the error between them.

\begin{lemma}\label{lem:q-net}
For an integer $L\ge2$, at most $C_0L^3$ values of $f$, each with error
at most $\delta$, determine an even real polynomial $p$ of degree at most
$L$ with
\begin{equation}\label{eq:q-sampled}
 \|f-p\|_\infty\le 5 b_L+4\delta.
\end{equation}
The constant $C_0$ is independent of $f,L,\delta$.
\end{lemma}

\begin{proof}
Put $h=(2\mathrm eL)^{-1}$ and choose a maximal $h$-separated set $X$.
Balls of radius $h$ cover the sphere. Those of radius $h/2$ are disjoint. Each has volume
$(2/\pi)\int_0^{h/2}\sin^2s\,ds\ge c h^3$. Thus $|X|\le C_0L^3$.

The restriction $T$ of a spherical polynomial of degree $L$ to a unit-speed
great circle is a trigonometric polynomial with frequencies from $-L$ to
$L$. For a complex argument $\xi+i\eta$ it satisfies
$|T(\xi+i\eta)|\le e^{L|\eta|}\max_{s\in\mathbb R}|T(s)|$.
For $\eta\ge0$ this follows from the maximum principle on the unit disk
applied to $e^{iLz}T(z)$ as a polynomial in $e^{iz}$.
For $\eta\le0$ use $e^{-iLz}T(z)$ and $e^{-iz}$.
Cauchy's derivative estimate on the circle of radius $1/L$ now gives
$\|T'\|_\infty\le \mathrm eL\|T\|_\infty$. Apply this bound along
the short arc from a point where \(|w|\) is largest to its nearest node.
Every spherical polynomial $w$ of degree at most $L$ satisfies
\begin{equation}\label{eq:q-norming}
 \|w\|_\infty\le 2\max_{x\in X}|w(x)|.
\end{equation}

Write \(y_x\) for the value returned at \(x\). Choose a real even
polynomial \(p\) of degree at most \(L\) satisfying
\(\max_{x\in X}|p(x)-y_x|\le b_L+\delta\).
These are finitely many linear inequalities in the polynomial coefficients.
They have a solution because \(v=T_{\lfloor L/2\rfloor+1}f\) satisfies
them. Appendix~\ref{app:value} gives a grid on which rational linear
feasibility suffices. For every feasible \(p\), the triangle inequality at
the nodes gives \(\max_X|p-v|\le2(b_L+\delta)\).
Equation~\eqref{eq:q-norming} now bounds
\(\|p-v\|_\infty\) by \(4(b_L+\delta)\). Adding
\(\|v-f\|_\infty\le b_L\) gives \eqref{eq:q-sampled}.
\end{proof}

Once $p$ is known, one can integrate $pt$ using polynomial arithmetic.
The smaller integral of $(f-p)t$ still has to be computed. A quantum
register has finitely many addresses, so we first replace this integral
by a weighted sum and estimate the error of that replacement.

For each integer $D\ge1$ fix a positive cubature rule $Q_D$ of total
mass one, exact on all spherical polynomials of degree at most $D$.
Here is one such rule in the coordinates $(u,\alpha,\beta)$ above.
Take $D+1$ equally spaced points in each angle, and a positive quadrature
on $[0,1]$ exact to degree $\lfloor D/2\rfloor$ in $u$. To obtain the
latter, use the zeros of the monic polynomial of degree
$r=\lfloor D/2\rfloor+1$ orthogonal to all lower degrees for $du$.
It has $r$ distinct interior zeros, since otherwise multiplication by the
polynomial on its interior sign changes would violate orthogonality.
Integrating the associated Lagrange interpolant defines a rule exact
through degree $2r-1$ by polynomial division and orthogonality.
Its weights are positive because each also integrates the square of
the corresponding cardinal polynomial. In the product rule a monomial
$z_1^a\overline z_1^{\,b}z_2^c\overline z_2^{\,d}$ of total degree
at most $D$ vanishes on angular averaging unless $a=b$ and $c=d$.
In the surviving case it is $u^a(1-u)^c$ of degree at most
$\lfloor D/2\rfloor$. The rule therefore has the asserted precision
and uses at most $(D+1)^3$ nodes.

Let $K\ge L$, and use $Q_D$ with $D=K+2$. The degree of $t$ is two.
If $v$ approximates $f$ within $b_K$ and has degree at most $K$, both
$pt$ and $vt$ are integrated exactly. Since $|t|\le1$ and both
probability measures have mass one,
\begin{equation}\label{eq:q-bias}
 \left|\int(f-p)t\,d\sigma-Q_D((f-p)t)\right|\le2b_K .
\end{equation}
Thus the fine mesh is selected by the desired error of integration,
independently of the number of queries used to take its mean.

The quantum step turns the weighted mean into the probability \(P\) that an auxiliary qubit is read as one. Write \(P=\sin^2\theta\). The prepared state has two components, distinguished by the value of this qubit. On their span, a pair of reflections acts as a rotation through \(2\theta\). Repeating this rotation coherently makes the angle accessible to a phase measurement, as in the amplitude-estimation algorithm of Brassard and coauthors~\cite[\S\,4]{BHMT}. With a budget proportional to \(M\), the resulting probability error is of order \(1/M\) at fixed confidence. For comparison, the sample mean of \(M\) independent readings of the auxiliary qubit has standard deviation \(\sqrt{P(1-P)/M}\). Here the input rotation \(U\) labels a sample point. The two reflections act on the computation registers used to estimate the sum.

\begin{theorem}\label{thm:q-finite}
Let $L\ge2$, $K\ge L$, $M\ge2$, and $\delta>0$, with $L,K,M$ integers.
There is an algorithm using at most $C_0L^3+C_1M$ queries of precision
$\delta$ which returns a number $\widehat a$. For each $f\in\mathcal F_*$,
with probability at least $8/\pi^2$ it satisfies
\begin{equation}\label{eq:q-finite}
 |\widehat a-a(f)|\le
 2b_K+\delta+
 2(5b_L+5\delta)\left(\frac{\pi}{M}+\frac{\pi^2}{M^2}\right).
\end{equation}
\end{theorem}

\begin{proof}
First use Lemma~\ref{lem:q-net}. Write $x_j,w_j$ for the nodes and
weights of the fine cubature, and $\widetilde f_j$ for the decoded oracle
values. Set $R=5b_L+5\delta$. Then
$s_j=(\widetilde f_j-p(x_j))t(x_j)$ belongs to $[-R,R]$.
Prepare the address register in the known state
$\sum_j\sqrt{w_j}|j\rangle$. With the value oracle and a rotation of an
ancillary qubit conditioned on $j$, encode $(1+s_j/R)/2$ as the
probability of its outcome $1$. Undo the computation of the value
register. The probability of this outcome on the entire state is
\begin{equation}\label{eq:amplitude-encoding}
 A=\frac12+\frac1{2R}\sum_jw_js_j .
\end{equation}
The amplitude-estimation theorem of Brassard, H{\o}yer, Mosca and Tapp
\cite[Theorem~12]{BHMT} uses $O(M)$ calls to this preparation and its
inverse. With probability at least $8/\pi^2$ the error in $A$ is at most
$2\pi\sqrt{A(1-A)}/M+\pi^2/M^2$.
Rescale by $2R$ and add the known coefficient of $p$.
The term $2b_K$ in \eqref{eq:q-finite} comes from \eqref{eq:q-bias}.
Rounding at the fine nodes costs at most $\delta$ in their weighted mean.
\end{proof}

For even \(L\) and a power of two \(M\), Appendix~\ref{app:value}
gives a binary implementation with at most \(4(10L+1)^3+4M-2\)
queries and an explicit allowance for the discretization of the mean.
For the query order, with \eqref{eq:q-classicalbound} choose
$K=\lceil (L+1)\sqrt M\rceil$ and $\delta=b_L/M$. Then the right side
of \eqref{eq:q-finite} is bounded by $C B (L+1)^{-2}/M$. A fixed positive
fraction of $N$ queries pays for the coarse net, and another fixed
fraction pays for the mean. Taking $L$ of order $N^{1/3}$ and $M$ of
order $N$ gives error at most $C' N^{-5/3}$. The fine cubature in this
choice can have order $N^{5/2}$ nodes. Its cost of coherent preparation
and the work of evaluating $p$ matter for circuit size or for emulation.

For an algorithm returning an ordinary real number, call $\varepsilon$
an admissible error when every input in $\mathcal F_*$ gives an output
within $\varepsilon$ of $a(f)$ with probability at least $3/4$.
Let $e_N^{\mathrm q}$ and $e_N^{\mathrm{ran}}$ be the infima of such errors
for quantum and classical randomized algorithms using at most $N$ function
queries on every run. Write $e_N^{\mathrm{det}}$ for the deterministic
worst-case error under the same budget. Points, digital precision and
postprocessing may be chosen as above. In the randomized model each
query observes the value at a single point.

Quantum integration on cubes was studied by Novak~\cite{Novak} and Heinrich~\cite{HeinrichSobolevQ}. For ordinary integration on \(S^3\), taking \(w=1\), \(r=2\), \(d=3\) and \(p=4\) in Li and Wang~\cite[Theorems~1.1(i) and~1.2(i)]{LiWang} gives the orders \(N^{-2/3}\) and \(N^{-7/6}\). Their randomized error is the worst expected absolute error and their sample budget is in expectation. In our fixed-budget problem Haar is the integration measure and \(t\) is part of the integrand. The finite exponent \(p=4\) gives a direct comparison for functions with bounded weak Laplacian.

Indeed, \eqref{eq:q-range} bounds \(\|f\|_\infty\) uniformly on the anchored class. The elliptic estimate of G\"uneysu and Pigola~\cite[Theorem~4.3]{GuneysuPigola} on the unit sphere gives
\begin{equation}\label{eq:elliptic-bound}
  \|f\|_{W^{2,4}(S^3)}
     \leq C\big(\|\mathcal L f\|_{L^4(S^3)}+\|f\|_{L^4(S^3)}\big).
\end{equation}
One may first smooth \(f\) by rotations and take a weak limit in \(W^{2,4}\). Multiplication by the smooth function \(t\) preserves this bound. To obtain the quantum upper exponent from Heinrich's theorem~\cite[Theorem~1, \(p=4\)]{HeinrichSobolevQ}, split \(tf\) by a fixed smooth partition of unity subordinate to coordinate cubes. Each summand, including the Haar density, has bounded \(W^{2,4}\) norm and compact support in the interior of its cube. Evaluating it requires only a value of \(f\) and known factors. The cubical rate is \(N^{-1-2/3}\) without a logarithmic factor. There are finitely many cubes, so the estimates can be added after a fixed number of repetitions to give the required success probability. Theorem~\ref{thm:q-finite} gives a group-based algorithm whose error depends explicitly on \(b_L\). The lower estimates below use the sign of \(t\) to construct perturbations which preserve both Haar mean zero and the anchoring condition.

The next theorem compares three kinds of algorithms for the coefficient \(a(f)\): deterministic, classical randomized and quantum. For each it gives the exact order of the best possible error in terms of the number \(N\) of values of the response that the algorithm may use.

\begin{theorem}\label{thm:q-rates}
For normalized Haar measure, the metric of the unit sphere and the coefficient
\eqref{eq:q-functional}, as $N$ tends to infinity,
\begin{equation}\label{eq:q-rates}
 e_N^{\mathrm{det}}\asymp N^{-2/3},\qquad
 e_N^{\mathrm{ran}}\asymp N^{-7/6},\qquad
 e_N^{\mathrm q}\asymp N^{-5/3}.
\end{equation}
Here the constants are independent of $N$. The same orders hold if the
anchoring condition $f(e)=0$ is replaced by $\int f\,d\sigma=0$.
\end{theorem}

\begin{proof}
The quantum upper estimate follows from Theorem~\ref{thm:q-finite}.
For the deterministic upper estimate use $a(p)$ after the coarse
sampling step. Its error is at most $5b_L+4\delta$. For the randomized
upper estimate draw $M$ independent fine nodes according to the cubature
weights. Each sampled $s_j$ lies in $[-R,R]$. The sample mean has variance
at most $R^2/M$, so with probability at least $3/4$ its error is at most
$2R/\sqrt M$. With the same $L,K,M,\delta$ as in the quantum construction,
\eqref{eq:q-bias} gives the asserted order. For the mean-zero class \(T_1f=0\) and \eqref{eq:even-first} bounds
\(\|f\|_\infty\) by \(h_1\). This supplies a fixed range for value
encoding in all three constructions.

We construct test functions whose coefficients count bits. Fix
\begin{equation}\label{eq:sign-flip-gate}
 J=\begin{pmatrix}0&1\\-1&0\end{pmatrix}.
\end{equation}
Left multiplication by $J$ preserves the metric of the unit sphere and Haar measure,
and $t(JU)=-t(U)$. Choose an open coordinate cube with compact closure
$V$ on which $t\ge t_0>0$. Choose it sufficiently small that the four
sets $V,-V,JV,-JV$ are disjoint and avoid $\{e,-e\}$. Such a choice
is possible near any point with $0<t<1$, since $J^2=-e$.
For each sufficiently large $k$ choose $k$ disjoint smaller coordinate
cubes of side $h$ inside $V$, with $c_1k^{-1/3}\le h\le c_2k^{-1/3}$.
This follows by subdividing a fixed smaller cube into
$\lceil k^{1/3}\rceil^3$ equal cubes.

In the $i$th small cube put a fixed smooth nonnegative interior bump,
translated and rescaled horizontally by $h$ and vertically by $h^2$,
and extend it by zero. Call the resulting function $v_i$.
In the fixed chart the coefficients of the second- and first-order
terms of $\Delta$ are bounded and the volume density is bounded above
and below by positive constants. Thus $\|\mathcal L v_i\|_\infty\le C$,
and $\int v_i\,d\sigma$ is between fixed positive multiples of $h^5$.
Make it even by setting $u_i(x)=v_i(x)+v_i(-x)$, and put
\begin{equation}\label{eq:paired-bump}
 w_i(x)=u_i(x)-u_i(J^{-1}x).
\end{equation}
The function $w_i$ is even, vanishes at $e$, and has mean zero. Its
positive and negative parts have opposite signs of $t$ on their
supports. In fact, $a(w_i)=2a(u_i)$, which is between fixed positive
multiples of $h^5$. The supports of different $w_i$ are disjoint.
Choose $A=\kappa h^5$ for a sufficiently small fixed $\kappa>0$ and set
$\psi_i=A w_i/a(w_i)$. Then $\|\mathcal L\psi_i\|_\infty\le1$ and
$a(\psi_i)=A$. For a bit string $b=(b_1,\ldots,b_k)$ define
\begin{equation}\label{eq:signed-test-function}
 f_b=\sum_{i=1}^k b_i\psi_i .
\end{equation}
Every $f_b$ belongs to both normalized classes under consideration,
and $a(f_b)=A\sum_i b_i$. At each point, its value depends on at most
one bit, whose index is determined by the point alone.

For a deterministic algorithm, run it on the zero input and set all
bits whose supports were queried equal to zero. Setting the unqueried
bits all equal to zero or all equal to one gives identical transcripts
and coefficients separated by at least $(k-N)A$. With $k=2N$ this forces
an error of order $N^{-2/3}$ on one of the two functions.

For a randomized algorithm let the bits be independent and uniformly
distributed. Allow each query to reveal the entire bit of its support.
Conditional on the algorithm's random seed and its transcript, the bits
still unobserved are independent fair bits. Their number $m$ is at least
$k-N$. The largest atom of their sum is
$2^{-m}\binom{m}{\lfloor m/2\rfloor}\le C/\sqrt m$, by Stirling's
formula. Set $k=2N$. Any interval of radius $cA\sqrt N$ can contain
at most $2c\sqrt N+2$ possible coefficient values. Choosing $c$ small
enough makes the conditional success probability less than $3/4$
for all sufficiently large $N$. The same holds after averaging over
seeds and observations. Some input therefore has error at least
$cA\sqrt N$, of order $N^{-7/6}$.

For a quantum algorithm each query \eqref{eq:q-oracle}, at any admissible
precision and on any finite address set, can be simulated by at most
two queries to the bit oracle for $b$. Compute the support index from the
address, read its bit coherently, compute and add the encoded value into
the value register, and undo the bit query. A known dummy zero bit
covers points outside all supports. If the coefficient is recovered
to error less than $A/2$, rounding its ratio to $A$ recovers the bit
count and hence its parity.

For completeness the bounded-error parity bound
\cite[Theorem~4.8 and Lemma~6.3 in the preprint]{Beals} follows
directly here. After $T$ bit queries each amplitude is a multilinear
polynomial of degree at most $T$ on the Boolean cube. A bit oracle raises
the degree by at most one and an input-independent unitary preserves the
bound. Measurements can be postponed by recording their outcomes in
ancillary registers. The expected sign of the reported parity is
therefore a polynomial of degree at most $2T$. Every monomial of degree
less than $k$ has zero average against $(-1)^{\sum_i b_i}$ on the
uniform Boolean cube. Success probability at least $3/4$ on every
input would instead give an average product of at least $1/2$.
Thus $2T\ge k$. Taking $k=4N+1$ rules out error $A/3$ with $N$ function
queries, since then $T\le2N$. Here $A\asymp N^{-5/3}$, giving the last
lower bound.
\end{proof}

\appendix
\section{Computing and checking a readout map}\label{app:computation}

For a fixed metric the approximation constant and its polynomial kernel can be computed once at each spin cutoff. The same kernel enters the shot estimator and its error calculation. All measures used for sampling the group have total mass one.

The certificate for \(h_k\) takes as inputs \(k\geq1\) and a desired enclosure width \(\xi>0\).
The outputs are a lower and an upper bound for \(h_k\) and a polynomial
kernel with a certified uniform approximation error on the unit class.
The following test works with approximate nodes. It certifies their
effect on the answer and does not require a validated nonlinear root solver.

Choose \(k\) distinct rational trial points \(0<v_1<\cdots<v_k<1\).
Write \(a_j=\arcsin\sqrt{v_j}\), \(a_0=0\), \(a_{k+1}=\pi/2\).
On \(a_i<a<a_{i+1}\) the sign \(s(\sin^2a)\) of
\(\prod_j(\sin^2a-v_j)\) is \((-1)^{k-i}\). Form the residual moments
\begin{equation}\label{eq:moment-residuals}
 m_r=\frac4\pi\sum_{i=0}^k(-1)^{k-i}
       \big[F_r(a_{i+1})-F_r(a_i)\big],\qquad 0\le r<k ,
\end{equation}
where \(F_r=A_r-A_{r+1}\) and
\begin{equation}\label{eq:sine-power-recurrence}
 A_0(a)=a,\qquad
 A_r(a)=\frac{2r-1}{2r}A_{r-1}(a)
          -\frac{\sin^{2r-1}a\cos a}{2r}\quad(r\ge1).
\end{equation}
Indeed \(F_r'(a)=\sin^{2r}a\cos^2a\). Interpolate the function \(K\) of
\eqref{eq:even-kernel} at \(v_j\), for instance by the Lagrange formula,
and write \(P(u)=\sum_{r=0}^{k-1}p_ru^r\).
The interpolation remainder has sign \(s\) by positivity of \(K^{(k)}\).
Equation~\eqref{eq:even-primitive} gives the integral norm
\begin{equation}\label{eq:app-upper}
 U=\int|K-P|\,d\nu
   =\frac2\pi\sum_{j=1}^{k}(-1)^{k-j}a_j\cos^2a_j
        -\sum_{r=0}^{k-1}p_r m_r .
\end{equation}

Set \(\mu_0=1\), \(\mu_{i+1}=\mu_i(i+1/2)/(i+2)\).
These are the moments of \eqref{eq:even-measure}, by the beta integral.
Solve the fixed rational system
\begin{equation}\label{eq:moment-correction}
 \sum_{j=0}^{k-1}\mu_{r+j}c_j=m_r,\quad 0\le r<k,
 \qquad d=\sum_{j=0}^{k-1}|c_j|.
\end{equation}
The matrix is positive definite since its quadratic form integrates a
polynomial squared against \(\nu\). The corrected function
\((s-\sum_j c_j u^j)/(1+d)\) has absolute value at most one and
annihilates all polynomials of degree below \(k\). Pair it with \(K-P\).
The contribution of \(s\) is \(U\) and the correction has magnitude
at most \(dU\). Hence
\begin{equation}\label{eq:app-enclosure}
    \max\{0,(1-d)/(1+d)\}\,U\ \le h_k\le\ U .
\end{equation}
In finite precision one encloses \(\pi\), the elementary functions, and
all subsequent operations by intervals with outward rounding. An exact
rational inverse of the moment matrix may be reused for every trial.
Use the larger of zero and the lower endpoint for \(U\)
and the upper endpoint for \(d\) on the left of \eqref{eq:app-enclosure},
and the upper endpoint for \(U\) on the right. Stop when the resulting
interval has width at most \(\xi\).

A convenient way to propose trial nodes is damped Newton iteration on
the angles, starting, for instance, from \(a_j=j\pi/(2k+2)\). The
entry in row \(r\), column \(j\) of its Jacobian is
\begin{equation}\label{eq:node-jacobian}
   -\frac8\pi(-1)^{k-j}\sin^{2r}a_j\cos^2a_j .
\end{equation}
Halve a step until the angles stay ordered and the residual decreases,
then round \(\sin^2a_j\) to distinct interior rational trial points.
Only \eqref{eq:app-enclosure} accepts a result. A slower guaranteed
alternative is to enumerate ordered dyadic nodes. At stage \(n\) test
the tuples of mesh \(2^{-n}\), refining the arithmetic until \(U\) and
\(d\) each have interval width at most \(2^{-n}\). Close to the true nodes
the interpolation points stay distinct, the moments tend to zero and
\(U\) tends to \(h_k\). Some stage therefore meets every positive
tolerance. The moment matrix becomes poorly conditioned as \(k\)
increases, so the arithmetic precision is part of this stopping rule.

For the first five cutoffs rational trial nodes with denominator \(10^{15}\) give enclosures by~\eqref{eq:app-enclosure} of width below \(2\cdot10^{-13}\), computed in 200-bit interval arithmetic with directed rounding. The nodes and the enclosures are listed in the code repository cited at the end of the paper. We write \(\bar h_k\) for the certified upper endpoints; to the digits of Table~\ref{tab:cutoffs} they agree with \(h_k\).

At a given cutoff a certified upper bound below the
target accepts it, and a certified lower bound above the target excludes
it. If a bracket straddles the target, one may refine it or test a
higher cutoff. Testing successive cutoffs with bracket width at most
half the target eventually accepts one, since \(h_k\to0\).

For a bound \(S\ge\|\mathcal Lf\|_\infty\), using \(P\) in
\eqref{eq:even-method} gives error at most \(SU\).
Rational coefficients \(\widehat p_r\) with
\(\sum_r|\widehat p_r-p_r|\leq\eta_0\) give error at most
\(S(U+\eta_0)\). If the data are the moments
\(\int y^\lambda f(y)\,d\sigma(y)\) through degree \(2k-2\), expand
the following kernel in monomials in \(y\) and sum its coefficients against the corresponding moments. The kernel is
\begin{equation}\label{eq:polynomial-density}
  1+\mathcal L_y\widehat P((x\cdot y)^2)
    =1-4u(1-u)\widehat P''(u)-(2-8u)\widehat P'(u),
    \qquad u=(x\cdot y)^2 .
\end{equation}
Indeed, for \(v=x\cdot y\) one has \(\Delta_y v=-3v\) and
\(|\nabla_y v|^2=1-v^2\). Applying the chain rule to \(u=v^2\)
gives the displayed coefficients.
If a supplied moment has absolute error \(\eta_\lambda\), add
\(\sum_\lambda |c_\lambda(x)|\eta_\lambda\) to the bound at \(x\),
where \(c_\lambda(x)\) is its coefficient in the expansion.
For a uniform bound replace \(|c_\lambda(x)|\) by the sum of the
absolute monomial coefficients of \(c_\lambda\), since \(|x_i|\le1\).

The number of readouts needed for a prescribed accuracy follows from~\eqref{eq:shot-map-uniform}. Let \(\varepsilon\) be the tolerance in positive-outcome probability, \(\alpha\) the failure probability and \(S_r\) the bound on the Laplacian. Choose \(k\) with \(S_rh_k<2\varepsilon\), form \(W_k\) by~\eqref{eq:polynomial-density}, and bound \(B_k\) and \(v_k\). With \(s=\gamma(2\varepsilon-S_rh_k)\) it suffices to take
\begin{equation}\label{eq:shots-budget}
 m\ge \frac{2(v_k+C_k s/3)}{s^2}\log\frac{2}{\alpha\omega}.
\end{equation}
For example, for \(S_r=25/3\) and \(\varepsilon=\alpha=0.05\) a certified kernel with \(k=11\) needs about \(2.4\) million readouts for a uniform guarantee.

To bound the largest probability error \(\mathcal P=\tfrac12\|r\|_\infty\), evaluate the fitted polynomial \(q\) of degree \(L=2k-2\) on a projective mesh of covering radius \(\rho\) with \(\mathrm eL\rho<1\). If \(A\) is the largest computed value, \(\zeta\) the error of one evaluation and \(E\) the bound for \(\|r-q\|_\infty\), the derivative estimate in the proof of Lemma~\ref{lem:q-net} gives
\begin{equation}\label{eq:peak-bracket}
 \frac{\max\{0,A-\zeta-E\}}{2}
 \le \mathcal P
 \le \min\!\left\{1,\frac{1}{2}\!
       \left(\frac{A+\zeta}{1-\mathrm e L\rho}+E\right)\right\}.
\end{equation}

Use a unit quaternion \(q=(q_0,q_1,q_2,q_3)\), with its scalar part first and the usual action by conjugation on imaginary quaternions. From three independent uniforms \(u,v,w\) in \([0,1)\), set \(\alpha=2\pi v\), \(\beta=2\pi w\) and put
\begin{equation}\label{eq:exp-map-quaternion}
 q=(\sqrt u\cos\alpha,\sqrt{1-u}\cos\beta,
        \sqrt{1-u}\sin\beta,\sqrt u\sin\alpha).
\end{equation}

The Haar moments of the monomials \(v^{2b}\) are
\begin{equation}\label{eq:exp-map-haar-moment}
          \int v^{2b}\,d\sigma(v)=
            \frac{\prod_{i=0}^3(1/2)_{b_i}}{(2)_{|b|}},
\end{equation}
where \((a)_j\) is the rising factorial. This follows by normalizing four independent real standard Gaussians to the unit sphere and integrating the radial variable.

Represent a gate by a unit quaternion, as above, so that \(U\cdot V\) is the Euclidean inner product in \(\R^4\). The four groups of Table~\ref{tab:polyhedral-designs} are the images of the following binary groups. The Pauli group consists of the eight quaternions \(\pm1,\pm\mathrm i,\pm\mathrm j,\pm\mathrm k\). The binary tetrahedral group adds the sixteen quaternions \((\pm1\pm\mathrm i\pm\mathrm j\pm\mathrm k)/2\). The binary octahedral group, which gives the Clifford gates, adds to these the twenty-four quaternions with two nonzero coordinates equal to \(\pm1/\sqrt2\). The binary icosahedral group adds to the binary tetrahedral group the ninety-six quaternions obtained from \((\pm\varphi,\pm1,\pm\varphi^{-1},0)/2\), with \(\varphi=(1+\sqrt5)/2\), by all sign choices and even permutations of the coordinates. Each set is closed under multiplication. Its design strength follows from the character averages of Section~\ref{sec:designs}. The first spins with a fixed vector are \(2,3,4,6\), respectively.

The value \(h_6=0.0147694\ldots\) was obtained from the node equations~\eqref{eq:moment-node-equations} in high-precision arithmetic. The errors for \(f_0\) are exact finite sums. For the Pauli gates the error is \(\mathrm e^{-2}/8\), since the four gates give \(t=1,1,-1,-1\).

The compression example of Section~\ref{sec:compression} uses the first-order Trotter steps
\[
 e^{-\mathrm i\Delta t\sum_iX_i}\,e^{-\mathrm i\Delta t\sum_iZ_iZ_{i+1}},
 \qquad \Delta t=1/40 .
\] The compressed layers have the same form with a separate angle for each \(ZZ\) bond and each \(X\) rotation, seven angles per layer. Starting from three plain Trotter steps, whose infidelity is \(4.4\cdot10^{-2}\), the BFGS method minimized \(1-|\operatorname{Tr}(\widetilde W^*W)|/16\). The responses were evaluated in a vectorized state-vector calculation. The grid is that of Appendix~\ref{app:value} with \(L=4\), and the local maximization started from \(400\) Haar points.

For the \(120\)-evaluation reconstruction, use the real projection kernel of \(\mathcal V_3\). In quaternion coordinates the addition formula of Section~\ref{sec:geometry} gives \(D(s)=80s^4-48s^2+3\), where \(s=x\cdot v\). Thus~\eqref{eq:design-reconstruct} gives \(d(x)=60^{-1}\sum_v d(v)D(x\cdot v)\) on the icosahedral nodes \(v\). For a fixed \(x\), the design integrates \(D(x\cdot v)^2\) exactly, and its Haar integral is the diagonal value \(D(1)=35\) of the projection kernel. If every sampled difference has absolute error at most \(\eta\), Cauchy--Schwarz therefore bounds the uniform reconstruction error by \(\sqrt{35}\,\eta\). If \(A\) is the computed maximum of the absolute reconstructed polynomial on the grid and each grid evaluation has absolute error at most \(\xi\), a uniform upper bound on the loss is \(2(A+\xi)+\sqrt{35}\,\eta\).

\section{A finite calculation with a numerical value oracle}\label{app:value}

The input and the output in this appendix are those of Section~\ref{sec:q-optimal}. The following budgets count rounded numerical evaluations of an anchored function and include calls used to undo an evaluation.

Suppose that \(f\) is real and even, with known \(S\geq\|\mathcal Lf\|_\infty\).
For \(S=0\), \(f\) is constant and \(a(f)=0\). Otherwise use a
representative \(g=(f-f(e))/S\), and multiply the final coefficient
and its error by \(S\). An oracle for \(g\) can also be supplied
directly. If the reference value is obtained by a separate query,
its error is included in the normalized value tolerance and the query
is counted once. For example, absolute errors at most \(S\delta/2\) in
each raw value \(f(x)\) and \(f(e)\) give a normalized difference
with error at most \(\delta\). The constant term of \(T_1 f\) is the
Haar mean. Applying its error bound at the identity and at a sample point
gives \(|g|\le 2h_1<1/2\).
In what follows \(\tau\) is the requested absolute coefficient error
in normalized units. For a requested error \(\varepsilon>0\) on \(a(f)\),
use \(\tau=\min\{1,\varepsilon/S\}\). Set
\begin{equation}\label{eq:app-design}
 b_n=\frac{3}{4(\lfloor n/2\rfloor+1)^2},\quad
 L=2\left\lceil\frac{\tau^{-1/5}}{2}\right\rceil,\quad
 \delta=\tau/8,\quad R_0=5(b_L+\delta).
\end{equation}
At degree \(n\) the first omitted integer spin is
\(k=\lfloor n/2\rfloor+1\). The sharp linear error \(h_k\) satisfies
\(h_k\le\beta_3/(4k^2)\), and \(\beta_3<3\).
A certified rational upper endpoint for \(h_k\) from
Section~\ref{app:computation} can also be used as \(b_n\).
Choose the smallest even \(K\ge L\) with \(2b_K\le\tau/4\), and
the first power of two \(M\ge2\) with
\(2R_0(\pi/M+\pi^2/M^2)\le\tau/4\).
For rational input all comparisons may use \(22/7\) instead of \(\pi\)
on the left. The total error of each normalized point value at its
specified node must be at most \(\delta\).

Use four charts for the coarse samples. Select a largest absolute
coordinate of a lift \(x\) and choose its sign positive. On its chart
write \(x=v/|v|\), with that coordinate of \(v\) equal to one and the
other three in \([-1,1]\). Take in each of the three coordinates all
values \(-1+2i/q\), \(i=0,\ldots,q\), where \(q=10L\).
A straight segment in the face has length at least that of its
normalized image, since \(|v|\ge1\) on the face. Thus the quotient
covering radius is at most \(\sqrt3/q\le1/(2\mathrm eL)\).
The proof of \eqref{eq:q-norming} applies to even polynomials, with at
most \(4(q+1)^3\) samples. Repeated projective nodes may be merged.

The basis consists of all monomials \(x^\lambda\) of total degree \(L\)
in four real coordinates. Any lower even degree can be raised to \(L\)
by multiplication by powers of \(\sum_i x_i^2=1\). On this grid the
entries \(v^\lambda/(v\cdot v)^{L/2}\) are rational.
For a rational tolerance and dyadic decoded data \(y_x\), solve the
rational linear feasibility problem
\begin{equation}\label{eq:coarse-fit}
      |p(x)-y_x|\le b_L+\delta\quad\text{at every coarse node}.
\end{equation}
Check every inequality on the returned coefficients. Feasibility follows
from the degree-\(L\) Favard operator, whose polynomial approximates
\(g\) within \(b_L\), together with the rounding allowance.
A rational solution exists for a nonempty rational polyhedron.
As in Lemma~\ref{lem:q-net}, a feasible \(p\) obeys
\(\|g-p\|_\infty\le5b_L+4\delta\). Thus the fine-node quantities
\((\widetilde g_j-p(x_j))t(x_j)\) lie in \([-R_0,R_0]\).
If the requested tolerance is given as a real number, a smaller positive
rational tolerance suffices throughout.

If the desired output is the entire approximating polynomial, one can
stop after this coarse step. For an absolute uniform tolerance
\(\varepsilon>0\), take the smallest even \(L\ge2\) with
\(5S b_L\le3\varepsilon/4\). Sample \(f\) directly with value error
at most \(\varepsilon/16\) and use \(Sb_L+\varepsilon/16\) as the
feasibility threshold. The same norming argument gives
\(\|f-p\|_\infty\le 5S b_L+4\varepsilon/16\le\varepsilon\).
The output is the list of rational monomial coefficients, valid on the
whole sphere. Rational upper bounds for \(S\) and a rational tolerance
from below suffice. This procedure uses point values. The smaller
sharp error in Section~\ref{app:computation} uses the harmonic moments.

Compute \(a(p)\) before querying the residual. Write
\(z_1=x_1+i x_2\) and \(z_2=x_3+i x_4\), so that
\(t=x_1^2+x_2^2-x_3^2-x_4^2\). A monomial with any odd exponent contributes
zero. For an even exponent \(2\nu\), the contribution per unit coefficient is
\begin{equation}\label{eq:app-monomial}
 a(x^{2\nu})=
 \frac{\prod_{i=1}^4(1/2)_{\nu_i}}{(2)_{|\nu|}}\,
 \frac{\nu_1+\nu_2-\nu_3-\nu_4}{|\nu|+2}.
\end{equation}
Here \((c)_r=c(c+1)\cdots(c+r-1)\), with \((c)_0=1\).
Integrating a Cartesian Gaussian monomial first by coordinates and then
in polar coordinates gives the first factor as the normalized spherical
moment of \(x^{2\nu}\). Adding one to each index in turn and inserting
the signs of \(t\) gives the second factor. Hence \(a(p)\) is rational.

To compute the finite mean, put \(D=K+2\), \(n=D+1\), \(r=\lfloor D/2\rfloor+1\), and \(J=rn^2\).
Let \(\lambda_1<\cdots<\lambda_r\) be the zeros of the Legendre
polynomial \(P_r\), with \(P_r(1)=1\), and set \(u_\ell=(1+\lambda_\ell)/2\).
In the coordinates of Section~\ref{sec:q-coefficient} take
\begin{equation}\label{eq:cubature-nodes}
  z_1=\sqrt{u_\ell}\exp(2\pi i h/n),\qquad
  z_2=\sqrt{1-u_\ell}\exp(2\pi i b/n),\qquad 0\le h,b<n .
\end{equation}
Order these addresses lexicographically by \((\ell,h,b)\).
At every such address \(t_j=\lambda_\ell\). The weight is
\begin{equation}\label{eq:cubature-weights}
 w_j=\frac{1}{n^2}\int_0^1
              \prod_{v\ne\ell}\frac{u-u_v}{u_\ell-u_v}\,du .
\end{equation}
This formula uses only real algebraic numbers and polynomial integration.
Equivalently, the usual Gauss--Legendre weights give
\(w_j=1/[n^2(1-\lambda_\ell^2)P_r'(\lambda_\ell)^2]\).
One can isolate the roots by rational bisection between consecutive
roots of \(P_{r-1}\), including the endpoints \(-1,1\).
The recurrence \(P_0=1,\ P_1(x)=x,\ 
(d+1)P_{d+1}=(2d+1)xP_d-dP_{d-1}\) supplies all the polynomials.
The cubature proof in Section~\ref{sec:q-optimal} gives the checks
\begin{equation}\label{eq:cubature-moments}
 w_j>0,\qquad \sum_j w_j=1,\qquad \sum_j w_jt_j=0,\qquad
 \sum_j w_jt_j^2=1/3 .
\end{equation}
Direct classical summation of the \(J\) decoded values gives
\(Q_D(\widetilde g\,t)\) with error at most \(2b_K+\delta\), before
arithmetic rounding. It provides a reference for the quantum estimate.

For a binary implementation let \(H,H'\) be the smallest powers of two
at least \(1\) satisfying \(H\ge16R_0J/\tau\), \(H'\ge16R_0/\tau\).
Index the nodes \(j=1,\ldots,J\), and form nondecreasing integers
\(B_0=0,\ldots,B_J=H\) with
\begin{equation}\label{eq:prefix-rounding}
  \left|B_j/H-\sum_{i=1}^j w_i\right|\le1/H .
\end{equation}
For example, enclose each cumulative weight to width \(1/(2H)\),
round \(H\) times its midpoint to the nearest integer, clip to
\([0,H]\), and take running maxima. Monotonicity of the exact cumulative
weights preserves the bound. Form uniform superpositions on the integers
\(v=0,\ldots,H-1\) and \(v'=0,\ldots,H'-1\) by Hadamard gates.
The interval \(B_{j-1}\le v<B_j\) determines the query address.
Decode the oracle value there and reversibly compute an integer
\(z_j\in[0,H']\) with
\begin{equation}\label{eq:amplitude-rounding}
 \left|\frac{z_j}{H'}-\frac{1+s_j/R_0}{2}\right|\le\frac1{H'},
 \qquad s_j=(\widetilde g_j-p(x_j))t_j .
\end{equation}
The rounded value is a dyadic number. Bounds on the known factors
\(p(x_j)\) and \(t_j\) may be computed in advance. Enclose the target
probability to width \(1/(2H')\), then round its
scaled midpoint and clip as above. Set a flag if \(v'<z_j\).
Reverse the arithmetic and the oracle call, retaining the flag and the
two uniform index registers. Denote this preparation by \(A\) and its
flag probability by \(P_d\). Unknown values are queried inside \(A\).
Building a classical table of all those values would itself require
\(J\) evaluations. The bias from the two binary encodings is at most
\begin{equation}\label{eq:discrete-mean-error}
 \left|R_0(2P_d-1)-\sum_j w_js_j\right|
 \le 2R_0J/H+2R_0/H' .
\end{equation}
Indeed, each discretized weight differs from \(w_j\) by at most \(2/H\),
and the discrete weights still sum to one.

Apply the phase-estimation circuit of \cite[Theorem~12]{BHMT} to
\(A|0\rangle\) with a control register of \(\log_2 M\) qubits.
For \(Q=-A S_0 A^\dagger S_1\), the reflections \(S_0,S_1\) reverse
the signs of the all-zero state and the flagged subspace, respectively.
Start the control register in a uniform superposition, apply the controlled
powers \(Q^{2^i}\) for \(0\le i<\log_2 M\), apply the inverse discrete
Fourier transform to the controls, and read \(y\in\{0,\ldots,M-1\}\).
The minus sign in \(Q\) contributes a \(Z\) on the control of the
first power. It cancels for every even power.
Return \(a(p)+R_0(2\sin^2(\pi y/M)-1)\), rounded with absolute error
at most \(\tau/8\) before restoring the original scale.
Each preparation or inverse uses two calls to \eqref{eq:q-oracle}.
The initial preparation and the \(M-1\) iterations of \(Q\) use at most
\(4M-2\) fine-node calls. A controlled reflection \(A S_0 A^\dagger\)
requires only the middle reflection to be controlled. State preparation,
lookup of the intervals, reversible arithmetic, and phase readout
also contribute to circuit size.

With probability at least \(8/\pi^2\) the absolute error in normalized
units is at most
\begin{equation}\label{eq:app-total}
  2b_K+\delta+
  2R_0\!\left(\frac{\pi}{M}+\frac{\pi^2}{M^2}\right)
   +\frac{2R_0J}{H}+\frac{2R_0}{H'}+\frac{\tau}{8}\ \le\ \tau .
\end{equation}
For an odd number \(s\) of independent readouts with the same coarse
data, return their median. If a single run succeeds with probability at
least \(p_0\), failure of the median has probability at most
\(\sum_{j=(s+1)/2}^{s}\binom{s}{j}(1-p_0)^j p_0^{s-j}\).
Here \(p_0=8/\pi^2\) for ideal gates. If the sum of operator-norm errors
of the gates in one complete run is at most \(\eta\), one may take
\(p_0=8/\pi^2-\eta\) for that circuit. Indeed, telescoping the unitary
product bounds the state-vector error by \(\eta\), and the variation
distance of the outcome distributions is bounded by the same number.
Median amplification requires \(p_0>1/2\).

Consider the normalized coefficient with \(S=1\), value precision
\(\delta=\tau/8\), and success probability \(p_0=8/\pi^2\).
A classical reference algorithm draws independent Haar rotations and
averages \(\widetilde g(U)t(U)\). Its bias is at most \(\delta/2\)
since \(\int|t|\,d\sigma=1/2\). By \eqref{eq:q-range} its variance
per draw is at most \((2\bar h_1+\delta)^2/3\), where \(\bar h_k\)
denotes the certified upper endpoint of Appendix~\ref{app:computation}.
Chebyshev's inequality gives a sufficient classical budget.
Together with the quantum construction above, the two counts are
\begin{equation}\label{eq:budget-comparison}
 N_{\rm MC}=\left\lceil\frac{(2\bar h_1+\delta)^2}
              {3(1-p_0)(\tau-\delta/2)^2}\right\rceil,\qquad
 Q=4(10L+1)^3+4M-2 .
\end{equation}
The first term of \(Q\) counts coarse point evaluations and the second
counts coherent calls, including uncomputation. Both algorithms receive
rounded numerical values of the same anchored function.

Table~\ref{tab:value-budgets} gives \(Q\) for two choices of its coarse
error bound. The simple choice is \(b_L=3/(4k^2)\) from
\eqref{eq:app-design}, with \(k=L/2+1\). The certified choice uses
\(b_L=\bar h_k\) for coarse reconstruction and for \(R_0\).
Both columns retain \(3/(4(K/2+1)^2)\) for the fine cubature bound and use
the smallest admissible power of two for \(M\). At \(\tau=1/256\)
the certified choice gives \(R_0<0.236108\) and \(M=2048\). The simple
choice gives \(R_0=2575/6144\) and \(M=4096\).
The coarse budget of \(275\,684\) evaluations is the same in both cases.

\begin{table}[ht]
\centering
\caption{Sufficient numerical-value queries at equal accuracy and confidence.
\(J\) is the number of addresses in the fine cubature. Values are rounded to two significant digits; the exact counts are computed by the script \texttt{budget.py} in the code repository.}\label{tab:value-budgets}
\begin{tabular}{@{}lrrrrr@{}}
\hline
\(\tau\)&\(L\)&\(J\)&\(Q\), simple&\(Q\), certified&\(N_{\rm MC}\)\\
\hline
\(1/256\)        &4&\(2.7\cdot10^5\)&\(2.9\cdot10^5\)&\(2.8\cdot10^5\)&\(2.6\cdot10^4\)\\
\(10^{-3}\)      &4&\(1.9\cdot10^6\)&\(3.4\cdot10^5\)&\(3.1\cdot10^5\)&\(3.9\cdot10^5\)\\
\(9\cdot10^{-4}\)&6&\(2.3\cdot10^6\)&\(9.4\cdot10^5\)&\(9.4\cdot10^5\)&\(4.9\cdot10^5\)\\
\(6\cdot10^{-4}\)&6&\(4.1\cdot10^6\)&\(9.7\cdot10^5\)&\(9.4\cdot10^5\)&\(1.1\cdot10^6\)\\
\(10^{-4}\)      &8&\(5.9\cdot10^7\)&\(2.4\cdot10^6\)&\(2.3\cdot10^6\)&\(3.9\cdot10^7\)\\
\hline
\end{tabular}
\end{table}

As the tolerance decreases, a jump in the coarse degree temporarily reverses the comparison.
On an interval where \(L,M\) are constant put \(q=Q\). Then
\(q<N_{\rm MC}\) holds exactly when \(\tau<\tau_q\), where
\begin{equation}\label{eq:budget-crossing}
 \tau_q=\frac{2\bar h_1}{(15/16)\sqrt{3(1-p_0)q}-1/8}.
\end{equation}
On \(10^{-4}\le\tau\le1\) the certified budget is smaller precisely
on \([1/1024,\,0.00113)\) and \([10^{-4},\,0.00065)\).
Just below \(1/1024\) the coarse degree changes from four to six and
the coarse budget increases to about \(9.1\cdot10^5\). For the simple choice
the corresponding upper endpoints are about \(0.00107\) and \(0.00064\).
These intervals can be checked by partitioning at \(\tau=L^{-5}\) for
even \(L\) and, for each fixed \(L\), at
\(\tau=40 b_L c_M/(1-5c_M)\), where
\(c_M=\pi/M+\pi^2/M^2<1/5\) and \(M\) ranges over powers of two.
The latter boundary is the amplitude error test above with equality.

The crossing also depends on how the classical samples are taken. Use the half-turn \(J\) of~\eqref{eq:sign-flip-gate}. It satisfies \(t(JU)=-t(U)\). Draw \(U\) with density \(2|t(U)|\) relative to Haar probability and use two queries per draw. For exact values the random variable
\begin{equation}\label{eq:paired-classical-value}
        Z=\tfrac14\operatorname{sgn}(t(U))\,[g(U)-g(JU)]
\end{equation}
has mean \(a(g)\) by left invariance and \(|Z|\le h_1/2\) by \(\|g-\bar g\|_\infty\le h_1\). To sample this distribution, choose \(t\) as a uniform sign times the square root of a uniform \([0,1]\) variate and choose the two phases as for Haar sampling. Couple the rounded and exact values at the same sampled commands. Rounding both queried values changes \(Z\) by at most \(\delta/2\). At success probability \(p_0=8/\pi^2\), Chebyshev's inequality gives a sufficient budget of \(2\lceil h_1^2/[4(1-p_0)(\tau-\delta/2)^2]\rceil\) calls. This is about \(9.7\cdot10^3\), \(1.5\cdot10^5\) and \(1.5\cdot10^7\) calls respectively at \(\tau=1/256,\,10^{-3},\,10^{-4}\). The classical sampling rule thus changes the ranking at an intermediate tolerance.

These comparisons concern sufficient budgets for the specified algorithms.
At \(10^{-4}\) the ratio of the Monte Carlo budget to the certified
quantum budget is about \(17.4\). With the paired classical rule the ratio is about \(6.5\).
Section~\ref{sec:q-optimal} also treats randomized integration that
exploits smoothness. The degree of parallelism, arithmetic and memory
costs of a classical implementation enter a time comparison, as do the
depth of reversible evaluation and the preparation of the cubature
weights in a quantum circuit. A sign readout estimates a point value
with standard error \(\sqrt{1-f(U)^2}/\sqrt m\) after \(m\)
independent repetitions. For variance near one, achieving pointwise
root-mean-square error \(\delta=\tau/8\) therefore costs about
\(6.4\cdot10^7\) repetitions at \(\tau=10^{-3}\) and
\(6.4\cdot10^9\) at \(\tau=10^{-4}\).
Those shots give a statistical estimate at one address. Equation~\eqref{eq:q-oracle}
requires reversible access to encoded numerical values in superposition.

For a known small circuit the coefficient can also be evaluated using its algebra. For example, \(a(f_\lambda)=0\) for every dephasing strength in \eqref{eq:dephasing-model}. At fixed \(t\), averaging over the phase of \(z\) gives mean \(t^2\), and \(t^3\) has Haar mean zero. The finite budgets above concern unknown numerical responses in the normalized class.

\medskip\noindent\emph{Methods and tools.} The numerical computations were carried out in Python with NumPy, SciPy and mpmath. The scripts that reproduce Tables~\ref{tab:cutoffs}--\ref{tab:state-constants} and~\ref{tab:value-budgets} are available at \url{https://github.com/alexeysolyanik/favard-su2}. Large language models were used as tools in drafting the text and the code. The author is responsible for all statements in the paper.

\end{document}